\documentclass[aps,prl,twocolumn,reprint, showpacs]{revtex4-1}
\pdfoutput=1
\usepackage[pdftex]{graphicx}
\usepackage{natbib}
\usepackage[cmex10]{amsmath}
\usepackage{color}
\usepackage{soul}
\usepackage{lipsum}

\begin{document}

%%%%%%%%%%%%%%%%%% title page information %%%%%%%%%%%%%%%%%%
\title{Discontinuous Electromagnetic Fields Using Huygens Sources For Wavefront Manipulation}

\author{Michael~Selvanayagam and George.~V.~Eleftheriades }
\affiliation{The Edward S. Rogers Department
of Electrical and Computer Engineering, University of Toronto, Toronto}

\date{\today}
%\email{gelefth@waves.utoronto.ca} %% email address is required



\begin{abstract}
We introduce the idea of discontinuous electric and magnetic fields at a boundary to design and shape wavefronts in an arbitrary manner. To create this discontinuity in the field we use electric and magnetic currents which act like a Huygens source to radiate the desired wavefront. These currents can be synthesized  either by an array of electric and magnetic dipoles or by a combined impedance and admittance surface. A dipole array is an active implementation to impose discontinuous fields while the impedance/admittance surface acts as a passive one.  We then expand on our previous work showing how electric and magnetic dipole arrays can be used to cloak an object demonstrating two novel cloaking schemes.  We also show how to arbitrarily refract a beam using a set of impedance and admittance surfaces. Refraction using the idea of discontinuous fields is shown to be a more general case of refraction using phase discontinuities. 
\end{abstract}


\maketitle
%%%%%%%%%%%%%%%%%%%%%%%%%%  body  %%%%%%%%%%%%%%%%%%%%%%%%%%
\section{Introduction}
One of the most basic boundary conditions in electromagnetic theory concerns the interface between two different materials. As propagating waves traverse the boundary they are reflected and refracted with different amplitudes.  This occurs specifically because the tangential electric and magnetic fields across the boundary are conserved.  One way to interpret this physical behaviour is that by maintaining the continuity of the tangential fields across a material boundary one can subsequently alter and control the propagation of electromagnetic fields in a region. This fact forms the basis for engineering and designing many electromagnetic and optical devices, including waveguides, lenses and scatterers. 
\\ \\
Because this idea is so fundamental to electromagnetic theory, the continuity of the fields across interfaces is often taken for granted. However it is worth asking a simple question. What if instead of using material interfaces, and thus continuous fields, to engineer desired wave propagation effects we could somehow impose discontinuities in the field, and thereby alter the propagation and amplitude of the wave. In this scenario, a desired discontinuity in the electromagnetic field is introduced across a boundary by purposely engineering the boundary itself. Here, we are building on work done in \cite{Selvanayagam_Eleftheriades_2012_4} where we introduced a discontinuous field at a boundary to cloak an object. We now seek to generalize this idea by exploring in more detail how such boundaries are realized, and how can they be used to alter and control the electromagnetic field in a variety of scenarios. 
\\ \\
Specifically, we approach this problem by proposing a general way to impose a discontinuity using currents at an interface. These currents will act like a Huygens source which radiates the desired wavefront to create the discontinuity in the electromagnetic field.  We first propose two ways to implement our discontinuity, one active and one passive. Using the active method we synthesize some novel cloaking examples building on our previously developed cloaking approach in \cite{Selvanayagam_Eleftheriades_2012_4}. With the passive method we develop a novel way to refract a plane-wave. Throughout these specific examples we will demonstrate how discontinuous fields can be used to create novel electromagnetic devices.

\section{Discontinuous Electromagnetic Fields}
\begin{figure}[!h]
\centering
\includegraphics[clip=true, trim= 0cm 0cm 0cm 0cm, scale=0.15]{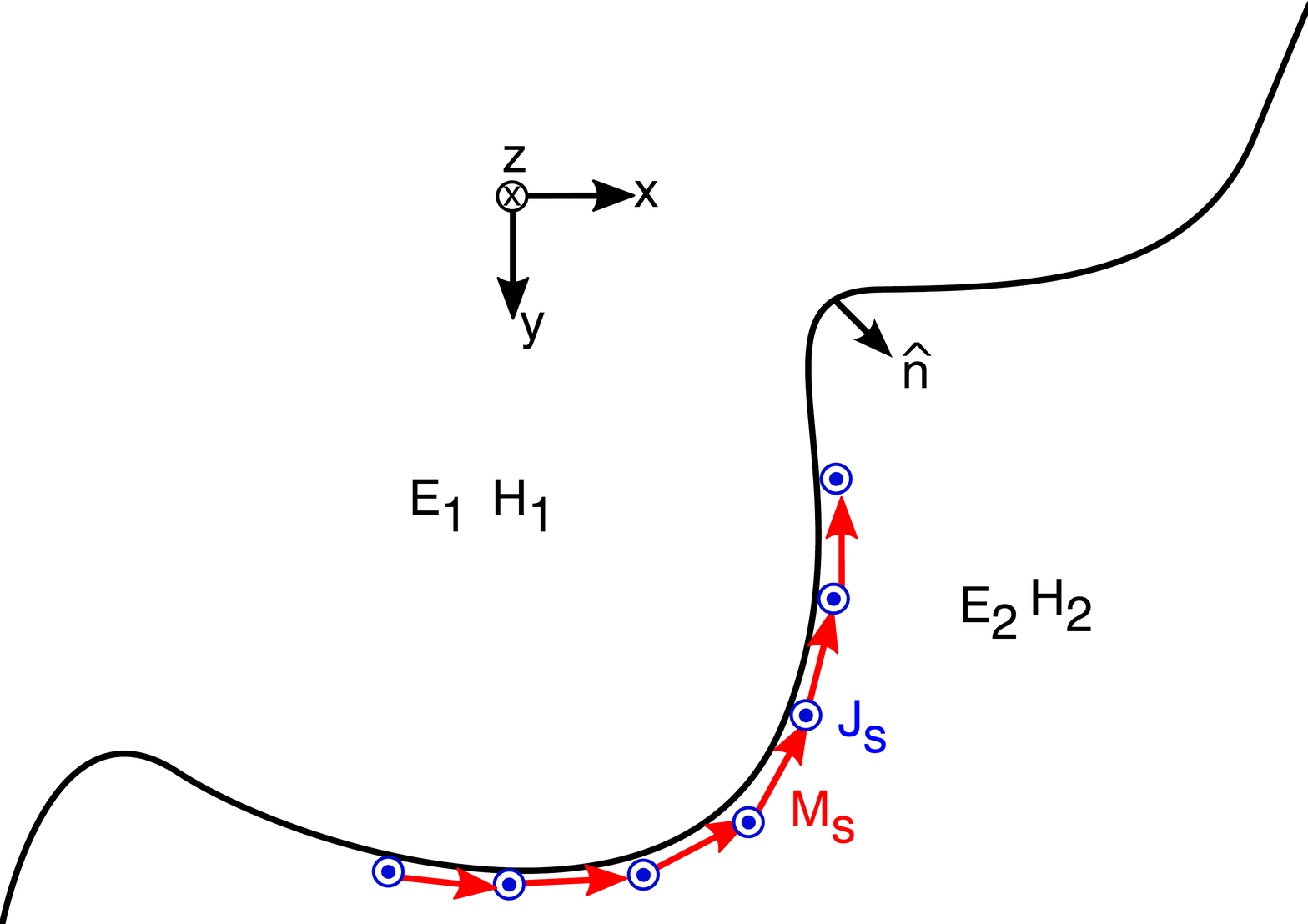}
\caption{An arbitrary boundary in free space upon which a discontinuity exists between the electric and magnetic fields on either side of the boundary. Electric and magnetic currents are imposed on the boundary to support the discontinuity. This is the motivating idea behind altering the fields using electric and magnetic currents.}
\label{fig:GeneralBoundaryDisc}
\end{figure}
The answer to the question of how one imposes a discontinuity in the electromagnetic field, follows from a simple examination of the boundary conditions presented in electromagnetic theory. To see this let us imagine an empty space (free space) without any materials present.  We then have an arbitrary boundary that divides our space into two halves and is described by a unit normal $\mathbf{\hat{n}}$ as shown in Fig.~\ref{fig:GeneralBoundaryDisc}. The two half spaces are also numbered respectively as shown. Without loss of generality we assume (throughout this paper) a field distribution that is invariant along the $z$-axis ($\frac{\partial}{\partial z}=0)$. We also assume a polarization of the fields with the electric field along the z-axis, $\mathbf{E}=E\mathbf{\hat{z}}$, and the magnetic field in the plane. 
\begin{figure}[!h]
\centering
\includegraphics[clip=true, trim= 3cm 3.5cm 2.5cm 2cm, scale=0.22]{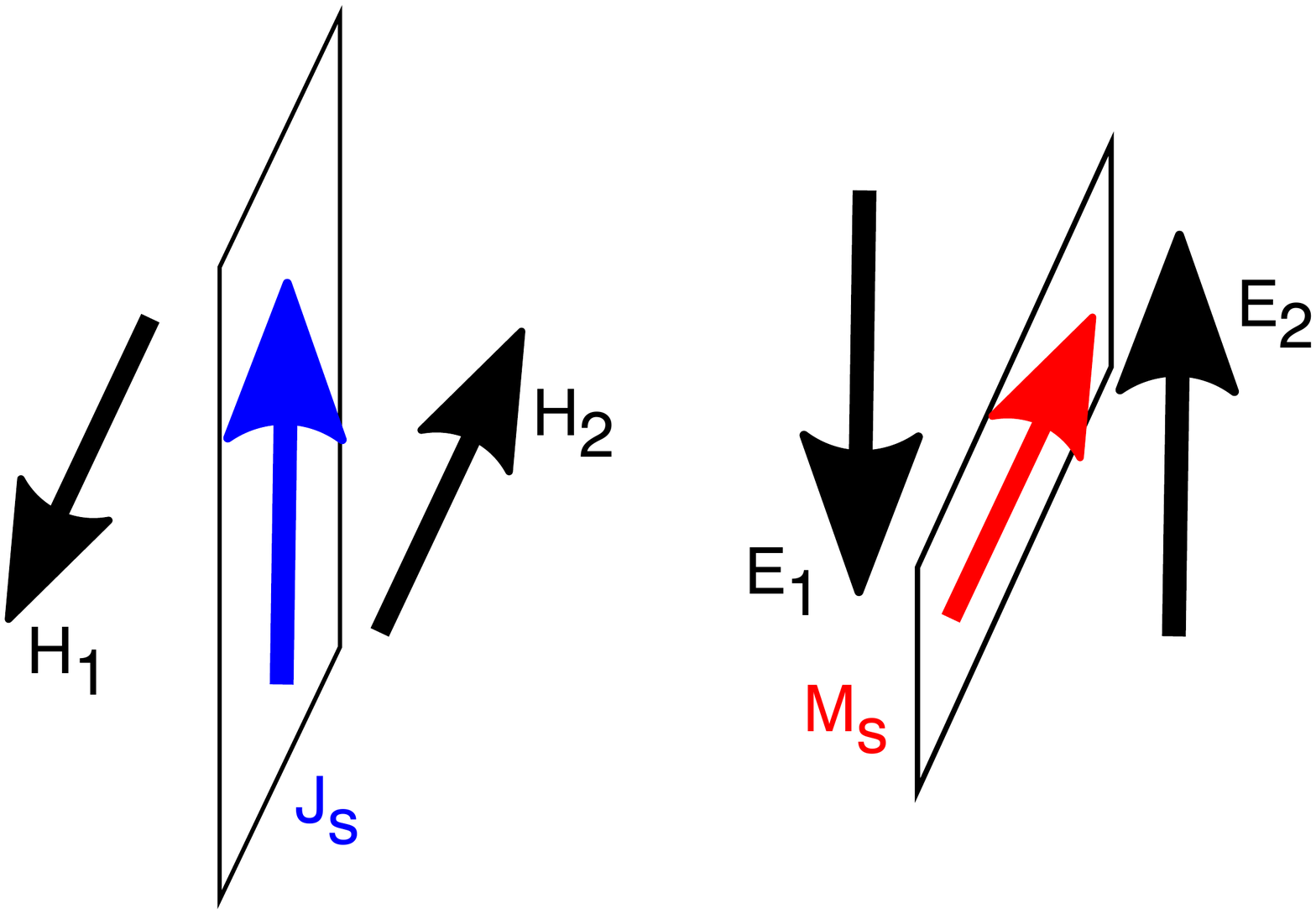}
\label{fig:CurrentDisc}
\caption{How currents on a surface impose a discontinuity.  Radiating electric currents create a magnetic field which curls around the current thus imposing a discontinuity in the magnetic field on either side of the boundary. Likewise, magnetic currents create an electric field which curls around the current imposing a discontinuity in the magnetic field}
\label{fig:CurrentDisc}
\end{figure}
\\ \\
On side one of the boundary we have some distribution of electromagnetic fields which we refer to as $\left\{\mathbf{E_1},\mathbf{H_1}\right\}$.  On side two of the boundary we have a different distribution of electromagnetic fields which are referred to as $\left\{\mathbf{E_2},\mathbf{H_2}\right\}$. Because the tangential fields are different on either side of the boundary we must place something at the boundary to satisfy the discontinuity since electromagnetic fields are naturally continuous.  From electromagnetic theory we find that what is required at the interface are electric and magnetic surface currents \cite{Harrington} equal to the discontinuity of the tangential fields. This is expressed as
\begin{eqnarray}
\label{eq:Ms}
\mathbf{\hat{n}}\times\left[\mathbf{E_2}-\mathbf{E_1}\right]&=&-\mathbf{M_s}, \\
\label{eq:Js}
\mathbf{\hat{n}}\times\left[\mathbf{H_2}-\mathbf{H_1}\right]&=&\mathbf{J_s}.
\end{eqnarray}
The presence of these surface currents enforce the discontinuity in the fields as shown in Fig.~\ref{fig:CurrentDisc}. Without the currents, the fields are naturally continuous and will not change their propagation characteristics. However, by placing these currents a discontinuity can be created \cite{Selvanayagam_Eleftheriades_2012_4}.
\\ \\
We now make a couple of observations about this scenario.  The first observation is the active nature of the boundary. By this we mean that these currents are impressed at the boundary and radiate a field to create the overall discontinuity in the field. As we will show below for specific examples, we can envision passive scenarios by constraining the amplitudes of the field on either side of the boundary. 
\\ \\
Another observation is that we require both an electric and a magnetic current at the boundary. This is because we need to alter both the electric and magnetic field across the interface to alter the propagation of the wave and thus we need two degrees of freedom, $\mathbf{J_s}$ and $\mathbf{M_s}$. This can be thought of as a Huygens source establishing the required discontinuity in the wave across the interface as it creates the desired wavefront. Also it is interesting to note that the discontinuity in the fields is enforced along a boundary only and does not require any bulk materials. A potentially beneficial result that will lead to thin conformal designs that we will further develop below. 
\\ \\
Finally, it can be noted that this idea of impressing electric and magnetic currents across an interface is not necessarily `new' as it has been used in theoretical and numerical electromagnetics when applying the equivalence principle \cite{Harrington}. However, what is different here is the interpretation of these currents as physical quantities which can be used to engineer the propagation of the electromagnetic wave across the boundary. This then raises another questions: how can these currents be physically realized?  To answer this we will introduce two ways to impress electric and magnetic currents along a boundary. Firstly, using active electric and magnetic dipoles. Secondly using passive impedance and admittance surfaces.

\subsection{Electric and Magnetic Dipoles}
\label{subsec:ElecMagDip}
The simplest way to implement a sheet of surface current, whether electric or magnetic, is to discretize the current using dipoles.  This is a well known idea from antenna array theory where discrete arrays are approximated by a surface current \cite{Hansen}. Here a set of electric and magnetic dipole arrays is needed. Such a configuration has been investigated for improving the receive characteristics of an antenna array \cite{Kwon_Pozar_2009}, though here we are looking at such combined arrays as a source for discontinuous wavefronts.  For the electric currents, the current is discretized by an array of electric dipole antennas. For the magnetic currents, magnetic dipoles can be used to discretize the currents. Because of the assumed polarization of the fields, the electric currents are along the $z$-axis, $\mathbf{J_s}=J_s\mathbf{\hat{z}}$ and the magnetic currents are in the plane and tangential to the boundary $\mathbf{M_s}=M_s\mathbf{\hat{t}}$, where $\mathbf{\hat{t}}$ is the unit tangent vector of the boundary. Because of this, the electric dipole antennas lie along the $z$-axis and the magnetic dipole antennas along the $\mathbf{\hat{t}}$ direction. 
\\ \\
The next question then is how many electric and magnetic dipoles are needed to appropriately discretize the continuous currents in Eqs.~\ref{eq:Ms} and \ref{eq:Js}. Assuming we can describe our boundary as a parameterized curve of coordinate $u$ , let us divide up the boundary into segments of length $d_u$ with a unit height $d_z=1$. Then the electric and magnetic dipole moments are given by $\mathbf{p_e}=j\omega\mathbf{J_s}d_ud_z$, $\mathbf{p_m}=j\omega\mathbf{M_s}d_ud_z$. These dipole moments can be approximated as point sources spaced every $d_u$ which gives, $\mathbf{p_{e,n}}=\mathbf{p_e}\delta(u-nd_u)$, $\mathbf{p_{e,n}}=\mathbf{p_e}\delta(u-nd_u)$.  The discretized electric and magnetic currents then are then given by a sum of these discrete dipole moments,
\begin{eqnarray}
\mathbf{J_{s,d}}=\sum_{n=-\infty}^{\infty}\mathbf{p_e}(n d_u)\delta(u-nd_u),\\
\mathbf{M_{s,d}}=\sum_{n=-\infty}^{\infty}\mathbf{p_m}(n d_u)\delta(u-nd_u).
\end{eqnarray}
The spacing of these dipole moments, $d_u$, must sample the currents so that they capture the `fastest' spatial variation of the currents at the boundary.  Thus from sampling theory \cite{Oppenheim}, we set $d_u$ to be small enough such that all the spatial variation of the electric and magnetic currents is captured. This can be found by taking the Fourier transform of $\mathbf{J_s}$ and $\mathbf{M_s}$. In some cases, $d_u$ may be different for the electric and magnetic currents leading to a different number of electric and magnetic dipoles.
\\ \\
%Another way to interpret these dipole moments is a surface of electric and magnetic polarizabilities.....
%\\ \\
To physically implement these electric and magnetic dipoles, the simplest way is to use small antennas. For the electric dipole, a traditional metallic strip of current can be used, while for the magnetic dipole a metallic loop of current would suffice.

\subsection{Impedance and Admittance Surfaces}
\label{subsec:ImpAdmSurf}
Another way to introduce a discontinuity in the fields is to use a surface which scatters the field.  Such surfaces can be characterized by an impedance \cite{Tretyakov}. These impedance surfaces form a boundary condition which have the following properties:
\begin{eqnarray}
\label{eq:EImpSurf}
\mathbf{\hat{n}}\times \left[ \mathbf{E_2}-\mathbf{E_1}\right] &=& 0, \\
\label{eq:HImpSurf}
\mathbf{\hat{n}}\times \left[ \mathbf{H_2}-\mathbf{H_1}\right] &=& \mathbf{J_s}, \\
\label{eq:ZImpSurf}
\mathbf{\hat{n}}\times\mathbf{E_1}&=&Z_s \mathbf{J_s}
\end{eqnarray}
where in general $Z_s$ is a complex impedance that relates the discontinuity in the magnetic field to the continuous electric field. Physically these surfaces are made up of planar metallic scatterers on which the induced currents radiate a scattered field. This scattered field when added with the incident field yields a total field which satisfies Eq.~\ref{eq:EImpSurf}- Eq.~\ref{eq:ZImpSurf}. 
\\ \\
Of course, given the discussion above, this boundary condition alone would not be able to introduce a discontinuity in the field because we need both electric and magnetic currents. Thus, arguing from duality \cite{Harrington}, we can envision another boundary condition which we will refer to as an admittance surface which is described by the following boundary condition:
\begin{eqnarray}
\label{eq:EAdmSurf}
\mathbf{\hat{n}}\times \left[ \mathbf{E_2}-\mathbf{E_1}\right] &=& -\mathbf{M_s}, \\
\label{eq:HAdmSurf}
\mathbf{\hat{n}}\times \left[ \mathbf{H_2}-\mathbf{H_1}\right] &=& 0, \\
\label{eq:YAdmSurf}
\mathbf{\hat{n}}\times\mathbf{H_1}&=&Y_s \mathbf{M_s},
\end{eqnarray} 
where $Y_s$ is a complex admittance that relates the discontinuity in the electric field to the continuous magnetic field. It is worth asking what such a surface would look like. In this case the scatterers would be metallic loop-like objects on which induced loops of current radiate a scattered field equivalent to what the magnetic current in Eq.~\ref{eq:EAdmSurf} would radiate. 
\\ \\
With these two boundary conditions, a combined surface made up of both impedance and admittance boundary conditions can be used to introduce a discontinuity in the electric and magnetic field.  To achieve this we can modify the scenario given in Fig.~\ref{fig:GeneralBoundaryDisc} by dividing up the fields on either side of the boundary. Here the electric and magnetic fields are divided into two groups. The first group is the continuous electric field and magnetic fields given by $\{\mathbf{E_{c1}}$, $\mathbf{E_{c2}}\}$ and $\{\mathbf{H_{c1}}$, $\mathbf{H_{c2}}\}$. At the boundary the continuous fields are equal to each other with $\mathbf{E_{c1}}=\mathbf{E_{c2}}$ and $\mathbf{H_{c1}}=\mathbf{H_{c2}}$.  The other group is the discontinuous fields which are given by $\{\mathbf{E_{d1}}$, $\mathbf{E_{d2}}\}$ and $\{\mathbf{H_{d1}}$, $\mathbf{H_{d2}}\}$. At the boundary these fields are not equal to each other.  Now by grouping the continuous electric field and the discontinuous magnetic field, an impedance at the boundary can be defined by applying Eq.~\ref{eq:EImpSurf}-Eq.~\ref{eq:ZImpSurf}.  Likewise, by grouping the continuous magnetic field and the discontinuous electric field an admittance can be defined from Eq.~\ref{eq:EAdmSurf}-Eq.~\ref{eq:YAdmSurf}. This combined surface can then be used to introduce a discontinuity in the electromagnetic field. 
\\ \\
In general without placing any constraints on the fields across the interface, $Z_s$ and $Y_s$ will be complex and the real part of both quantities could be less than zero \cite{Grbic_Merlin_2008}. An impedance or admittance with a negative real part implies that the surface is active and supplying power to the system. However as we will show below we can envision specific scenarios where the real part is constrained to be passive and lossless. 

%Do we introduce transmission-line model

\section{Enforcing Discontinuities to Cloak an Object} 
As we demonstrated in \cite{Selvanayagam_Eleftheriades_2012_4} an array of electric and magnetic dipoles can be used to cloak an object. Using this  idea of discontinuous electromagnetic fields we introduce two other cloaking schemes which can hide an object by introducing a discontinuity in the field at the boundary.  
\\ \\
In the first example, a discontinuity is imposed on the boundary of a scatterer illuminated by a plane-wave.  By carefully choosing this discontinuity we can create a cloak which cancels the scattered field outside the cloak while leaving the fields undisturbed inside the cloak. This creates a cloak which can interact with the field inside the volume without disturbing the field outside, similar to the anti-cloak proposed in \cite{Chen_etal_2008,gallina_engheta_2010} but without using bulk materials.
\\ \\
\begin{figure}[!h]
\centering
\includegraphics[clip=true, trim= 0cm 4cm 0cm 3cm, scale=0.25]{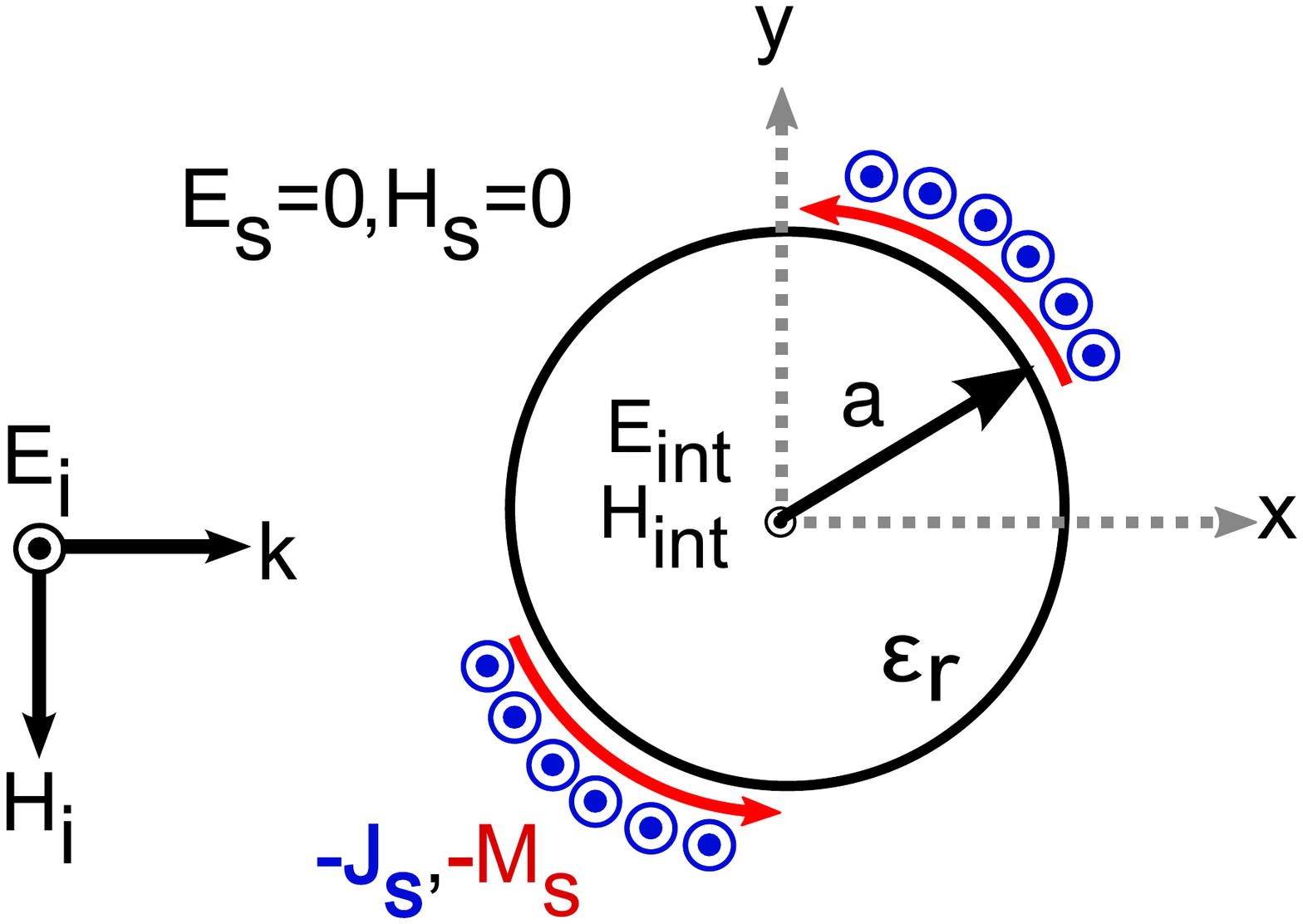}
\caption{A schematic of the active cloak which is constructed by enforcing electric and magnetic dipoles on the boundary of the dielectric cylinder. In this section two cloaking schemes are demonstrated by altering the weights of the electric and magnetic dipoles on the boundary.  In the first example the fields are canceled outside of the scatterer without disturbing the interior fields $\{\mathbf{E_{int}},\mathbf{H_{int}}\}$. In the second example the electric and magnetic dipoles create a scattered field exterior to the cloak that looks like a metallic cylinder.}
\label{fig:ActiveCloakSchem}
\end{figure}
Our object to be cloaked is a dielectric cylindrical scatterer situated at the origin and of infinite extent along the $z$-axis. The radius of the scatterer is $a=0.7\lambda$ at an operating frequency of 1.5~GHz while the dielectric constant of the material is $\epsilon_r=10$. The incident field is a plane-wave described by $E_i e^{-jkx}\mathbf{\hat{z}}$. This is depicted in Fig.~\ref{fig:ActiveCloakSchem}. The total fields outside the scatterer can be decomposed into an incident field and a scattered field, while the fields inside the scatterer are left in terms of the total field.  At the boundary of the scatterer these fields are continuous. However we want to impose a discontinuity in the field at the boundary of the scatterer by adding the negative of the scattered field outside the object without disturbing the fields inside the scatterer. This gives a discontinuity that is described in terms of the negative of the scattered field, $-\mathbf{E_s}$,$-\mathbf{H_s}$. By plugging the scattered field into Eq.~\ref{eq:Ms}-\ref{eq:Js}, a set of electric and magnetic currents are found which are given by,
\begin{eqnarray}
\label{eq:MsCloak1}
\mathbf{M_s}=-\mathbf{n}\times(-\mathbf{E_s}) &= \nonumber \\-k^2\sum_{n=-\infty}^{n=\infty}A_{sn} H_n^{(2)}(k\rho)e^{-jn(\phi+\frac{\pi}{2})}\mathbf{\hat{\phi}},\\
\label{eq:JsCloak1}
\mathbf{J_s}=\mathbf{n}\times(-\mathbf{H_s})&= \nonumber \\ j\omega\epsilon k \sum_{n=-\infty}^{n=\infty}A_{sn} H_n^{(2)'}(k\rho)e^{-jn(\phi+\frac{\pi}{2})} \mathbf{\hat{z}},
\end{eqnarray}
where $A_{sn}$ are the coefficients of the scattered field expansion in terms of cylindrical wave functions and are given in \cite{Harrington}. By imposing these electric and magnetic currents at the boundary of the object, the negative of the scattered field is radiated creating a discontinuous field at the boundary which leaves no scattered field outside while the fields inside the scatterer are undisturbed. Following Section.~\ref{subsec:ElecMagDip}, the currents are implemented using electric and magnetic dipoles. To sufficiently sample the currents the Fourier transform of $M_s$ and $J_s$ is taken and it is found that $24$ electric and magnetic dipoles placed along the boundary of the of the scatterer are sufficient. Generally, the larger the scatterer, the greater the number of dipoles that are required to sufficiently sample the electric and magnetic currents.
\\
\\
\begin{figure*}[!t]
\centering
\includegraphics[clip=true, trim= 2cm 11.0cm 1cm 4cm, scale=0.5]{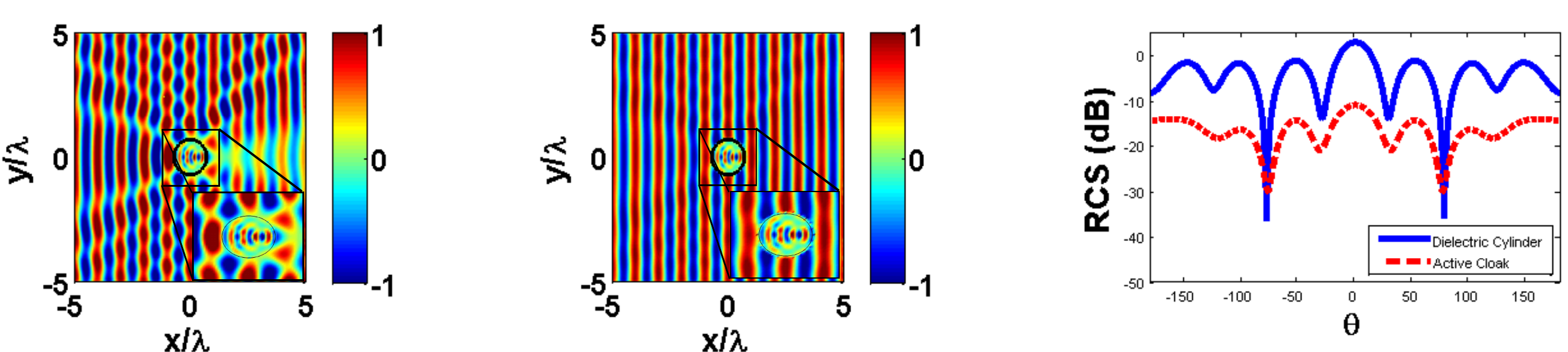}
\label{fig:DielCylScatteredCloakInset}
\caption{An active cloak which cancels the scattered field without disturbing the fields inside the dielectric. On the left is a plot of the total out-of-plane electric field, $E_z$ for the bare dielectric cylinder without a cloak. The inset shows the fields inside the cylinder. The dielectric boundary is marked with a black circle. The middle plot shows the cylinder with a cloak made up of electric and magnetic dipoles.  We can see that the field pattern now resembles the incident plane-wave only. Note also that the fields in the dielectric are relatively undisturbed as shown in the inset. Finally the 2D radar cross section (RCS) is shown on the right indicating a decrease in the scattering off of the cylinder. }
\label{fig:DielCylScatteredCloakInset}
\end{figure*}
We can simulate this array of electric and magnetic dipoles surrounding a dielectric cylinder using COMSOL multiphysics.  The results are shown in Fig.~\ref{fig:DielCylScatteredCloakInset} where the total field of the bare  dielectric cylinder is shown along with the total field for the cylinder surrounded by electric and magnetic dipoles. Also plotted is the 2D bistatic radar cross section of the cylinder defined in two-dimensions as $\sigma_{RCS}=2\pi r\frac{|E_s|^2}{|E_i^2|}$ \cite{Gibson}. We can see that the scattering off of the cylinder is reduced while the fields inside the dielectric are undisturbed as shown in the inset.  Such a scenario would allow for interesting applications in sensing where any disturbance in the field could be minimized while still being detected.
\\ \\
The second example of a discontinuity imposed along the boundary of a scatterer further extends this idea of canceling the scattered fields without disturbing the fields inside.  Here, along with canceling the scattered fields of the object without disturbing the fields inside, we further superimpose a set of fields outside the scatterer which are the scattered fields of a different object. This makes the dielectric scatterer look like a different object. In this specific example we make the dielectric cylinder look like a perfectly conducting cylinder. To achieve this we must impose a discontinuity in the field through a set of electric and magnetic currents which are given by ,
\begin{eqnarray}
\label{eq:MsCloak2}
\mathbf{M_s}=-\mathbf{n}\times(-\mathbf{E_s+E_{sm}}) &= \nonumber \\ \Big[ -k^2\sum_{n=-\infty}^{n=\infty}A_{sn} H_n^{(2)}(k\rho)e^{-jn(\phi+\frac{\pi}{2})} + \nonumber \\ k^2\sum_{n=-\infty}^{n=\infty}A_{smn} H_n^{(2)}(k\rho)e^{-jn(\phi+\frac{\pi}{2})} \Big] \mathbf{\hat{\phi}},\\
\label{eq:JsCloak2}
\mathbf{J_s}=\mathbf{n}\times(-\mathbf{H_s+H_{sm}}) &= \nonumber \\ j\omega\epsilon k \Big[ \sum_{n=-\infty}^{n=\infty}A_{sn} H_n^{(2)'}(k\rho)e^{-jn(\phi+\frac{\pi}{2})}- \nonumber \\ \sum_{n=-\infty}^{n=\infty}A_{smn} H_n^{(2)'}(k\rho)e^{-jn(\phi+\frac{\pi}{2})} \Big]\mathbf{\hat{z}},
\end{eqnarray}
where $\mathbf{E_{sm}}$ and $\mathbf{H_{sm}}$ are the scattered fields of a perfectly conducting cylinder of the same radius as the dielectric scatterer and 
$A_{smn}$ are the scattering coefficients of the perfectly conducting cylinder, given by $A_{smn}=-J_n(ka)/(k^2 H_n^{2}(ka))$ \cite{Harrington}. These electric and magnetic currents create a field outside of the cylinder which looks like the field of a perfectly conducting cylinder while canceling the scattered field of the dielectric object along without disturbing the fields inside. To implement the currents, electric and magnetic dipoles are used again and like the previous example, $24$ dipoles of either kind are required to sufficiently sample the currents for the same geometry.
\\ \\
\begin{figure*}[!t]
\centering
\includegraphics[clip=true, trim= 4cm 9cm 0cm 2cm, scale=0.45]{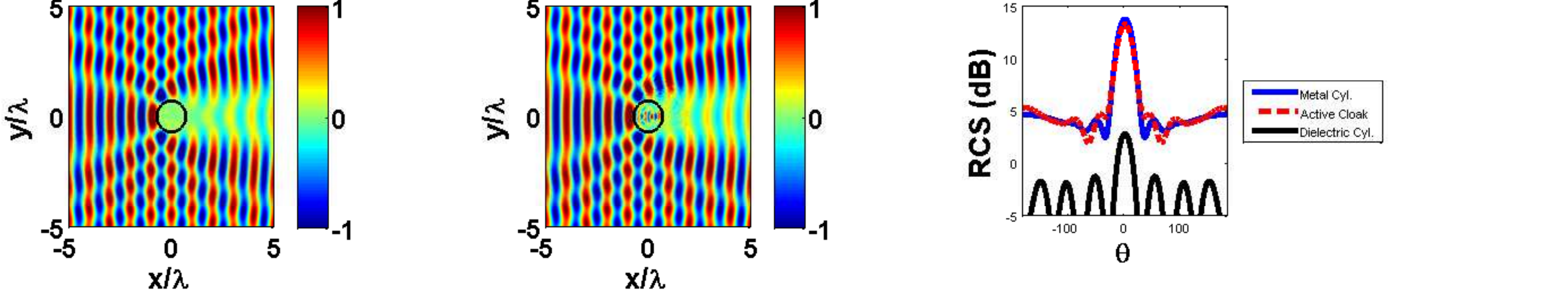}
\label{fig:DielCylScatteredCloakInset}
\caption{An active cloak which cancels the scattered field of the dielectric cylinder while making the dielectric cylinder look like a metallic cylinder. On the left is a plot of the total out-of-plane electric field, $E_z$ for a bare metallic cylinder. The boundary is marked with a black circle and is the same size as our dielectric cylinder. The middle plot shows the dielectric cylinder with a cloak made up of electric and magnetic dipoles.  We can see that the field pattern now resembles the total field of our metallic cylinder, which disguises the dielectric cylinder. Note also that the fields in the dielectric are undisturbed as well. Finally the 2D radar cross section is shown on the right demonstrating quantitatively how the dielectric cylinder resembles a metallic cylinder when cloaked with the active dipole array. }
\label{fig:DielCylScatteredSpoofCloak}
\end{figure*}
This configuration is also simulated using COMSOL multiphysics and the results are shown in Fig.~\ref{fig:DielCylScatteredSpoofCloak}. Here the total field of a bare perfectly conducting cylinder is shown along with the total field of the active cloak. Here we can see from the electric field pattern, we have made a dielectric cylinder look like a conducting cylinder.  Again the field inside the dielectric cylinder is undisturbed.  Finally the 2D bistatic radar cross sections is shown and we can demonstrate a good agreement between the scattering of a metal cylinder and our cloak and the noticeable difference between a bare dielectric cylinder. This demonstrates the ability of this array of electric and magnetic dipoles to disguise a dielectric cylinder as a metallic cylinder by imposing a discontinuity in the field at the boundary.

\subsection{Comments}
With these basic examples it becomes clear that being able to impose a discontinuity in the electromagnetic field is a powerful concept which allows for the realization of some novel applications. We can see that by imposing electric and magnetic currents at a boundary through discrete dipole arrays, the wavefronts can be manipulated in many possible ways.  As we have demonstrated so far by having both electric and magnetic dipoles we are able to radiate a wavefront which creates our desired field as per Huygens principle.  
\\ \\
A distinguishing feature of this proposed approach stems from the need to know the scattered field and thus the incident field \emph{a priori}. This is inherent in this approach of creating discontinuous fields, as the discontinuity must be inserted into a known field distribution \cite{Miller_2006}. However for applications where the incident field is known or can be determined using other methods, this approach shows significant promise.

\section{Enforcing Discontinuities to Refract and Reflect A Plane Wave}
Another interesting example of wavefront manipulation is through altering the direction of a plane wave either through refraction or reflection. As stated earlier, this is typically done with bulk materials which alter the fields by forcing continuity across the boundaries. Here we now apply the idea of creating a discontinuity in the electromagnetic field to refract a plane wave.  
\\ \\
\begin{figure}[!h]
\centering
\includegraphics[clip=true, trim= 0cm 4cm 0cm 3cm, scale=0.25]{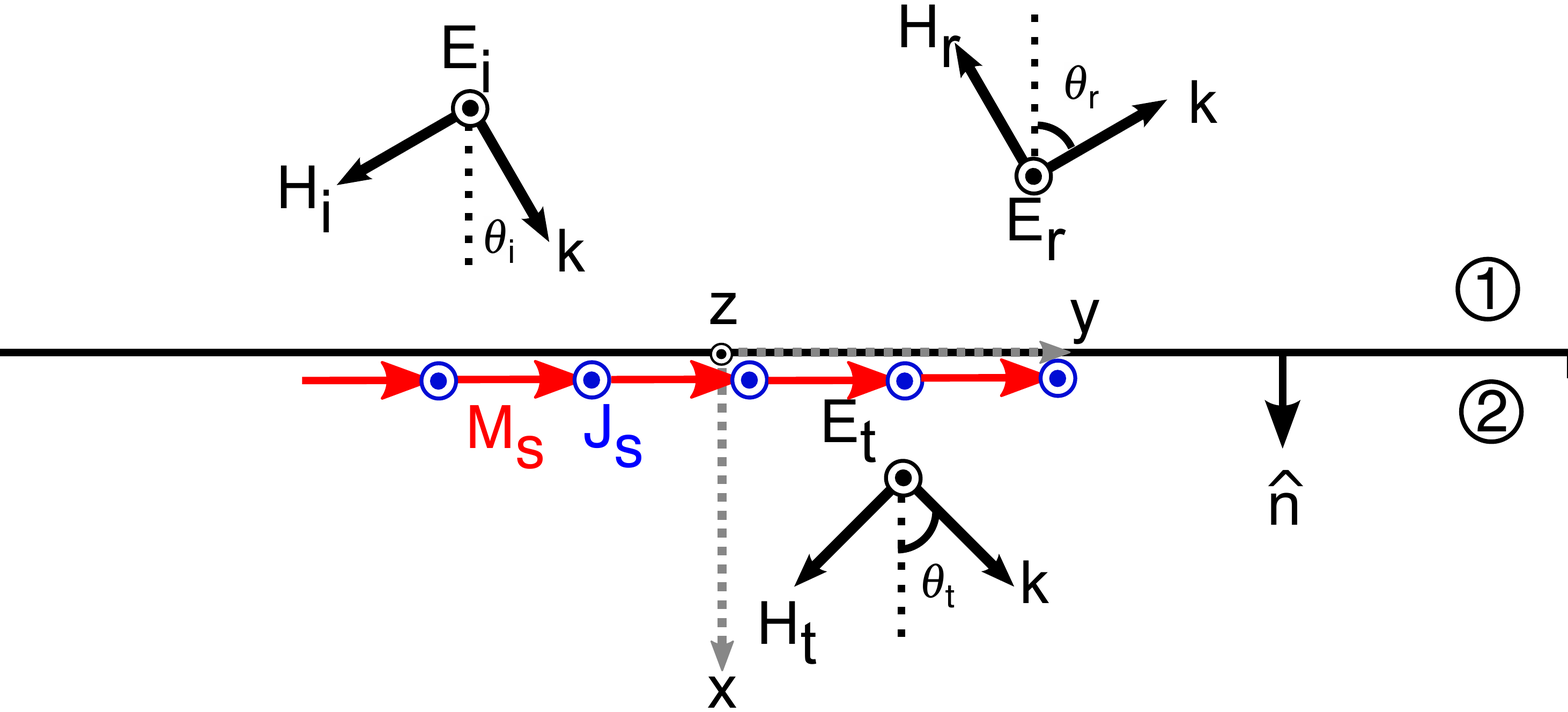}
\caption{Refraction  and reflection at an interface. If the incident plane-wave is refracted or reflected at the boundary in free-space, a discontinuity exists at the interface which requires an electric and magnetic current at the interface.  We can impose those currents using a discrete active electric and magnetic dipole array or we can use a passive impedance and admittance surface, provided we constrain the amplitudes of the fields. }
\label{fig:PlaneWaveRefractionActive}
\end{figure}
The basic scenario is described in Fig.~\ref{fig:PlaneWaveRefractionActive} where a plane wave in free-space is incident on a boundary located at $x=0$. At this boundary, a discontinuity in the field is imposed and the wave suddenly bends as it traverses the boundary.  To design the boundary we will look at two ways to impose the discontinuity using active electric and magnetic dipoles as well as a passive impedance/admittance surface.

\subsection{Refraction Using Active Arrays of Electric And Magnetic Dipoles}
\label{subsec:ActiveRefraction}
To use an array of electric and magnetic dipoles to refract a plane-wave wave we can simply subtract the differences between the incident and refracted plane-wave at the surface. Using Eqns.~\ref{eq:Ms} and  \ref{eq:Js} we can find the electric and magnetic currents which are given by
\begin{eqnarray}
\mathbf{M_s}&=&-\mathbf{\hat{y}} \left[ E_t e^{-jkx\cos\theta_t} - E_i e^{-jkx\cos\theta_i} \right], \\
\mathbf{J_s}&=&-\mathbf{\hat{z}} \left[\frac{1}{\eta} \cos\theta_t E_t e^{-jkx\cos\theta_t} - \frac{1}{\eta} \cos\theta_i E_i e^{-jkx\cos\theta_i} \right],
\end{eqnarray}
where $E_i$ and $\theta_i$ are the amplitude and direction of the incident wave and $E_t$ and $\theta_t$ are the amplitude and direction of the refracted wave and $\eta=\sqrt{\frac{\mu_o}{\varepsilon_o}}$ is the impedance of free-space.  Here we have assumed that there is no reflected wave generated by the interface.  We also note a similar solution could be found for a reflected wave only. 
Following again the procedure given in Section~\ref{subsec:ElecMagDip}, an array of electric and magnetic dipoles can be constructed at the $x=0$ plane. Here we need to place an electric and magnetic dipole every $2\lambda/3$ to sufficiently sample the continuous electric and magnetic currents.  
\\ \\
\begin{figure}[!h]
\centering
\includegraphics[clip=true, trim= 1cm 6cm 2cm 7cm, scale=0.2]{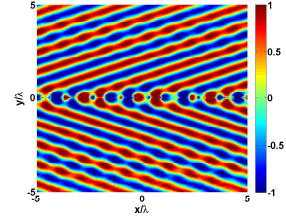}
\label{fig:ActiveNegativeRefraction}
\caption{An array of electric and magnetic dipoles at a surface which interfere to generate a negatively refracted plane-wave. The total electric field along the vertical axis is plotted. The incident field is along the $\theta_i=20^{\circ}$ direction and the refracted wave along the $\theta_t=-20^{\circ}$.  }
\label{fig:ActiveNegativeRefraction}
\end{figure}
Again we can simulate this scenario using COMSOL multiphysics. In our specific example we generate negative refraction where a plane-wave incident at an angle of $\theta_i=20^{\circ}$ refracts to an angle of $\theta_t=-20^{\circ}$.  This is depicted in Fig.~\ref{fig:ActiveNegativeRefraction}. This demonstrates the basic concept of a discontinuous electromagnetic field at a planar interface.  Again we can see how both electric and magnetic currents are required to radiate the refracted wavefront.  
\\ \\
For refraction, it is worth asking if this is possible using a passive method like in `normal' refraction. Thus we will investigate the use of impedance and admittance surfaces to refract an incident plane-wave. To ensure the passivity of the impedances we will look at constraining the amplitudes of the refracted fields.

\subsection{Other Ways Of Refracting A  Plane Wave At A Surface}
It is worth pausing here to discuss other methods of refracting a plane wave using a surface as this is a very active area in the literature. In the microwave regime, transmitarrays and reflectarrays have been used as scattering surfaces to bend or reflect incident waves into arbitrary directions as they pass through a surface \cite{Ryan_etal_2010,Lau_Hum_2012}. These surfaces can often be characterized as impedance surfaces \cite{Lau_Hum_2012}. For these microwave devices multiple surfaces are often used to optimize bandwidth and/or to minimize reflections at the surface, however the main concept can be reduced to a single surface. At the infrared and optical wavelengths similar designs have been constructed using either passive arrays of nano-antennas \cite{Yu_etal_2011,Kildishev_etal_2013} or blazed grating-like structures \cite{Tsai_etal_2011}.
\\ \\
Despite the different language used between the microwave and optical communities, these ideas all follow a similar logic that is easily described using a refraction law derived from a phase discontinuity between the refracted and incident fields \cite{Yu_etal_2011,Steyskal_etal_1979}. Here a linear phase shift along the elements that make up the surface are used to alter the direction of the incident plane-wave. This is expressed as 
\begin{equation}
\label{eq:GeneralRefraction}
\sin\theta_t-\sin\theta_i=\frac{1}{k_o}\frac{d\Phi}{d y},
\end{equation}
where $\Phi=ky(\sin\theta_t-\sin\theta_i)$ is the phase difference along the surface between the incident and refracted wave. Thus if we have a linear phase variation along a surface, we expect an incident plane-wave to refract along the $\theta_t$ direction after passing through the surface. This idea follows somewhat 	analogously from linear antenna array theory \cite{balanis}. However, as demonstrated in \cite{Lau_Hum_2012}, the problem with structures that provide a linear phase-shift along a surface is that they always excite other modes (other plane-waves or Floquet modes for periodic structures) which propagate in different directions other then the desired angle of refraction $\theta_t$. And if one examines the simulated or measured results in the papers listed above, this behaviour can be observed. 
\\ \\
There are two changes which can be made to these designs to get around this problem. The first and most fundamental change is to impose both electric and magnetic currents to create a proper discontinuity in the electromagnetic field as we have proposed.  The second change is that the elements which make up our scattering surfaces do not vary linearly to impose a linear phase shift in the electromagnetic field. We will now demonstrate this using both impedance and admittance surfaces to refract a plane-wave.

\subsection{Refraction Using Passive Impedance And Admittance Surfaces}
As stated, impedance and admittance surfaces can be used to implement a discontinuity in the electromagnetic field.  In line with the previous work described in the last section we would like to make our impedance and admittance surfaces passive and lossless, implying $\Re(Z_s) = 0$ and $\Re(Y_s)=0$. This then requires us to carefully construct the amplitudes of the waves which scatter off of the surface.  To find the required $Z_s$ and $Y_s$ to refract a plane-wave, we will break up the fields into a set of continuous electric fields and discontinuous magnetic fields and vice versa as stated in Section.~\ref{subsec:ImpAdmSurf}.  
\\ \\
For both the impedance and admittance surfaces illuminated by a plane-wave we can describe a set of fields which scatter off of the surface and satisfy the boundary conditions given in Eq.~\ref{eq:EImpSurf}-Eq.~\ref{eq:YAdmSurf}.  Because of the symmetry across the $y$-axis, both the impedance and admittance surfaces scatter the incident wave into plane-waves which propagate from either side of the boundary in the same direction. This forces us to analyze the problem by constructing a symmetric set of scattered fields on either side of the surface.
\begin{figure}[!h]
\centering
\includegraphics[clip=true, trim= 0cm 4cm 0cm 3cm, scale=0.25]{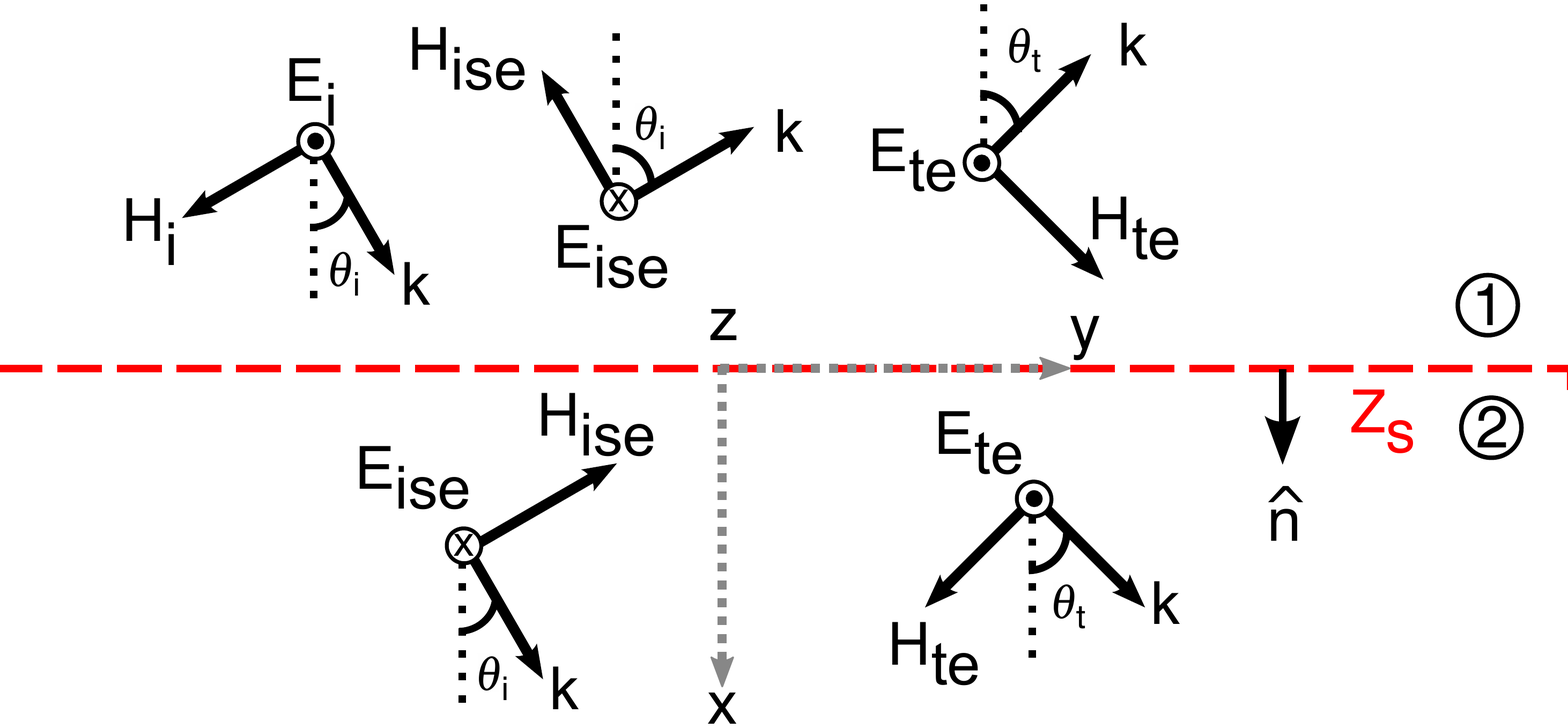}
\caption{A plane-wave incident on an impedance surface. Shown are the incident and the scattered fields.  On side 2 of the boundary the total fields are the scattered fields summed with the incident plane wave (not-shown). }
\label{fig:PlaneWaveRefractionImpSurf}
\end{figure}
\begin{figure}[!h]
\centering
\includegraphics[clip=true, trim= 0cm 4cm 0cm 3cm, scale=0.25]{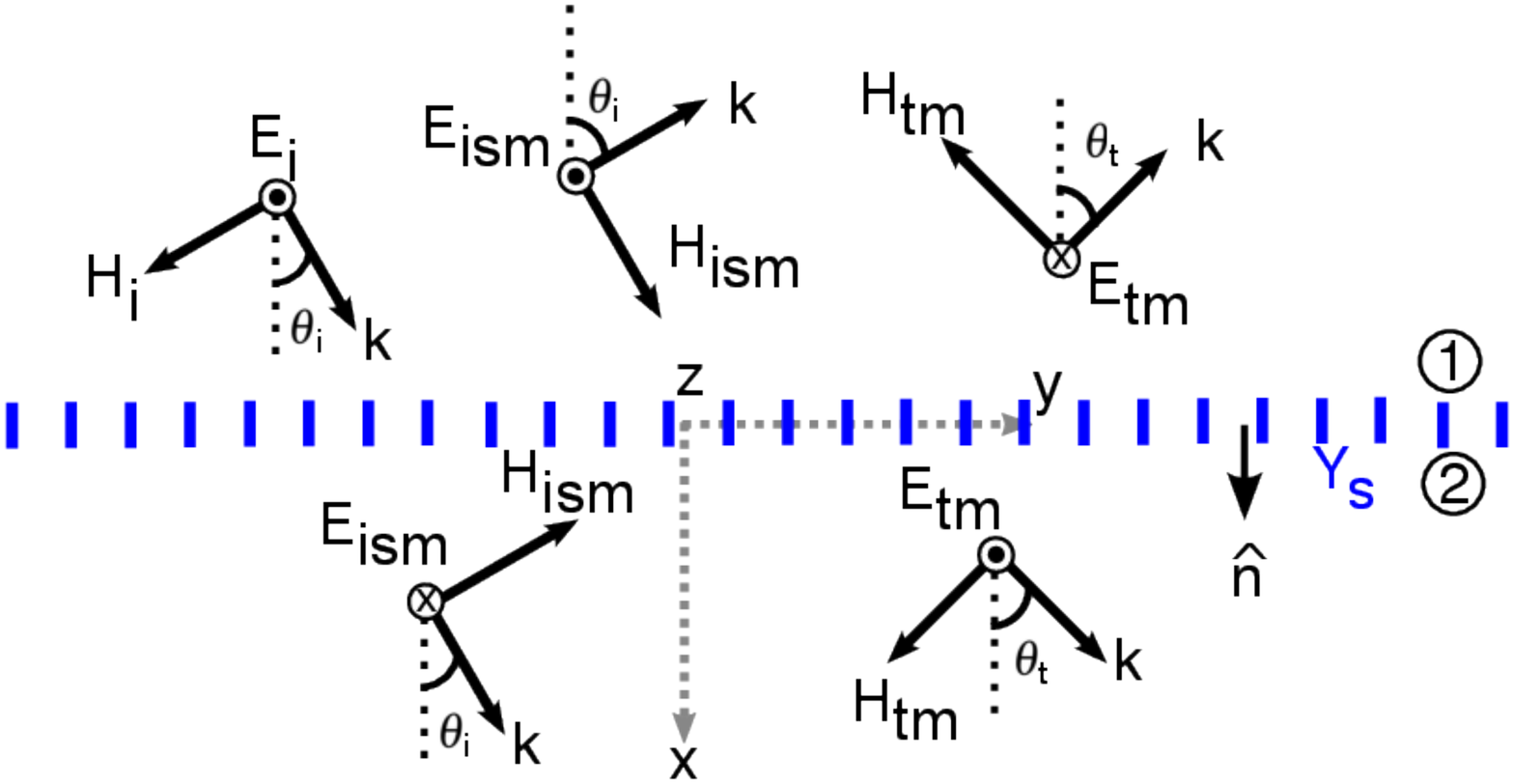}
\caption{A plane-wave incident on an admittance surface. Shown are the incident and the scattered fields.  On side 2 of the boundary the total fields are the scattered fields summed with the incident plane wave (not-shown). To create a surface which primarily reflects the incident plane-wave as opposed to refracting it, the sign of $E_{tm}$ should be flipped.}
\label{fig:PlaneWaveRefractionAdmSurf}
\end{figure}
\\ \\
Recalling that our ultimate goal is to arrive at a total field which refracts the incident plane-wave across the boundary, our combined impedance and admittance surface has to do two things. First it must cancel the incident field that exists beyond the interface on side two. Second it must scatter the field into the refracted beam on side two as well.  Thus an incident plane-wave scattering off of either the impedance or admittance surface must generate four-scattered plane-waves, two on either side of the interface. This is pictured in Fig.~\ref{fig:PlaneWaveRefractionImpSurf} for the impedance surface and in Fig.~\ref{fig:PlaneWaveRefractionAdmSurf} for the admittance surface. These scattered fields in Fig.~\ref{fig:PlaneWaveRefractionImpSurf} form a set of continuous electric fields and discontinuous magnetic fields which will define the impedance surface. The scattered fields in Fig.~\ref{fig:PlaneWaveRefractionAdmSurf} form a set of continuous magnetic fields and discontinuous electric fields which correspondingly define the admittance surface.
\\ \\ 
We can now use these fields to analyze the impedance and admittance boundary condition respectively. Starting with the impedance boundary condition we plug these fields shown in Fig.~\ref{fig:PlaneWaveRefractionImpSurf} into Eq.~\ref{eq:ZImpSurf} to give us
\begin{widetext}
%\[
\begin{eqnarray}
E_i e^{-jkx\sin\theta_i} - E_{ise} e^{-jkx\sin\theta_i}  +E_{te} e^{-jkx\sin\theta_t} = \nonumber \\   Z_s \Big[  
-\frac{1}{\eta}\cos\theta_i E_i e^{-jkx\sin\theta_i} +\frac{1}{\eta}\cos\theta_i E_{ise} e^{-jkx\sin\theta_i} -\frac{1}{\eta}\cos\theta_t E_{te}  e^{-jkx\sin\theta_t} \nonumber \\  +\frac{1}{\eta}\cos\theta_i E_i e^{-jkx\sin\theta_i} +\frac{1}{\eta}\cos\theta_i E_{ise} e^{-jkx\sin\theta_i} -\frac{1}{\eta}\cos\theta_t E_{te} e^{-jkx\sin\theta_t}  \Big],
\end{eqnarray}
%\]
\end{widetext}
We can now solve for $E_{ise}$ and $E_{te}$, which are the amplitudes of the scattered fields, by forcing $\Re(Z_s)=0$.  This will also give us an expression for $\Im(Z_s)=X_s$. Doing this we find that
\begin{eqnarray}
E_{ise}=\frac{E_i \cos\theta_t}{\cos\theta_i+\cos\theta_t} \\
E_{te}=\frac{E_i \cos\theta_i}{\cos\theta_i+\cos\theta_t}
\end{eqnarray}
And our expression for the surface reactance is,
\begin{eqnarray}
\label{eq:Xs}
X_s=-\frac{\eta (\cos\theta_t+\cos\theta_i)}{4\cos\theta_t\cos\theta_i}\cot(\Phi/2),
\end{eqnarray}
where $\Phi=ky(\sin\theta_t-\sin\theta_i)$, the phase shift between the refracted and incident plane waves.  Note that our expression for $X_s$ is not linear with respect to $y$ (or $\Phi)$, implying that we need a non-linear gradient in our impedance surface to generate a linear phase-shift for our refracted wave. Also note that this is the best possible case for a single impedance surface as we can channel the incident plane-wave into a minimum of four other plane-waves, and thus only succeed in partially refracting the beam. 
\\ \\
We now do the same for the admittance surface at $x=0$. Taking the fields in  Fig.~\ref{fig:PlaneWaveRefractionAdmSurf} and inserting them into Eq.~\ref{eq:YAdmSurf} and forcing $\Re(Y_S)=0$ we get
\begin{eqnarray}
E_{ism}=E_{tm}=\frac{E_i \cos\theta_i}{\cos\theta_i+\cos\theta_t},
\end{eqnarray}
Again we also get an expression for the surface susceptance,
\begin{eqnarray}
\label{eq:Bs}
B_s=-\frac{\cos\theta_t}{2 \eta}\cot(\Phi/2)
\end{eqnarray}
which is a non-linear function of $y$, the spatial coordinate along the boundary. Note also that both Eq.~\ref{eq:Xs} and Eq.~\ref{eq:Bs} depend on $\Phi$, the phase shift between the incident and refracted wave.
\\ \\
\begin{figure}[!h]
\centering
\includegraphics[clip=true, trim= 0cm 4cm 0cm 3cm, scale=0.25]{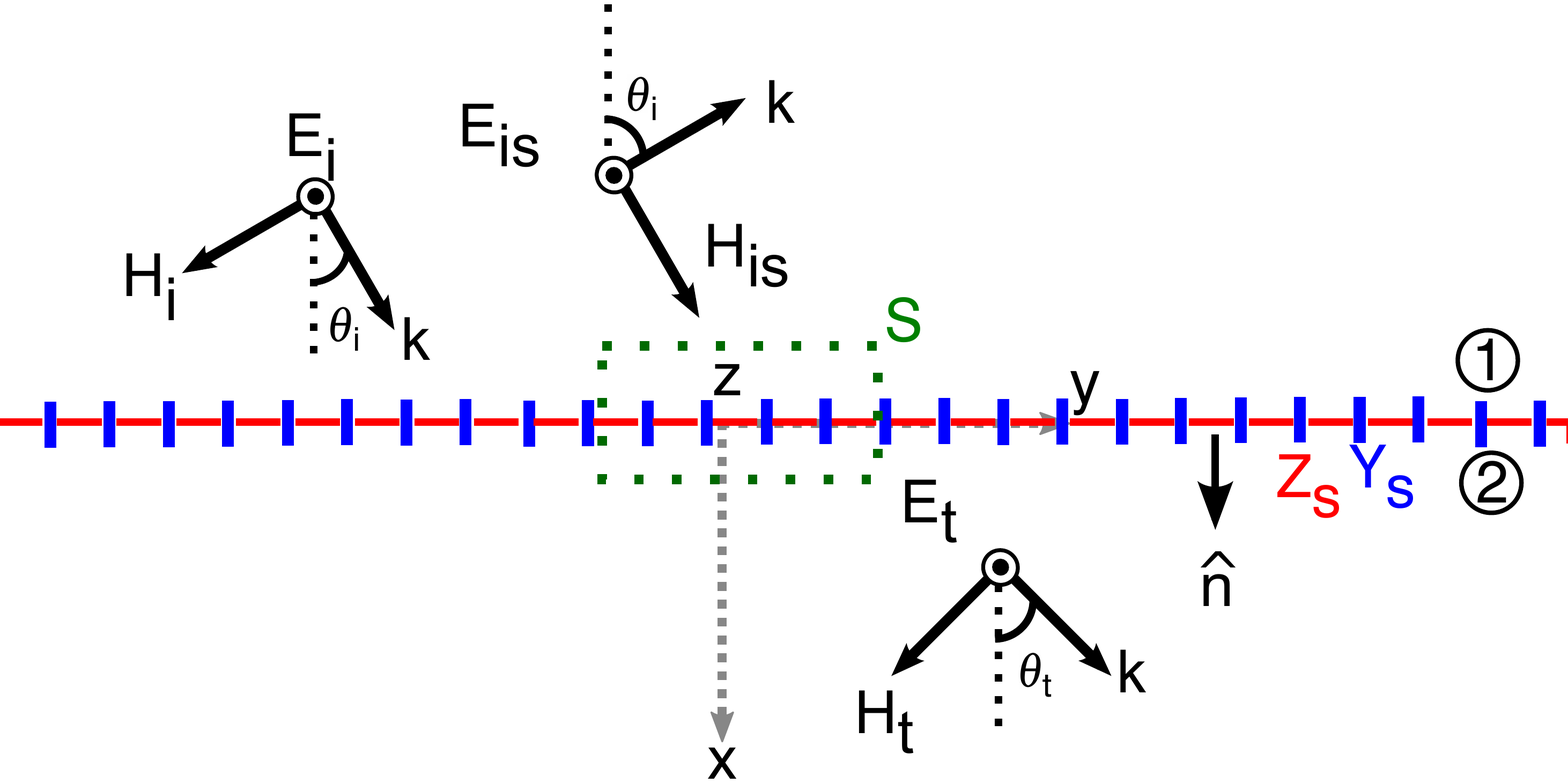}
\caption{A plane-wave incident on a combined impedance and admittance surface. Shown are the incident and the total fields on either side of the boundary. Because the impedance and admittance surface create induced electric and magnetic currents we can successfully refract the incident plane-wave without any other fields on side two of the boundary. Also note that the majority of the power is scattered into $\mathbf{E_t}$ as the amplitude of $\mathbf{E_{is}}$ is much smaller as shown in Fig.~\ref{fig:SurfaceRefractionDesignCurves} }
\label{fig:PlaneWaveRefractionImpAdmSurf}
\end{figure}
As we have stated before, the complete effect comes when our impedance and admittance surfaces are superimposed creating a discontinuity in both the electric and magnetic field. Superimposing the scattered fields off of both the impedance and admittance surface we get the following set of fields shown in Fig.~\ref{fig:PlaneWaveRefractionImpAdmSurf}.  We now have an incident plane-wave which when scattering off of both admittance and impedance surfaces gives us a refracted field with an amplitude given by,
\begin{eqnarray}
\label{eq:RefractedAmp}
E_{t}=E_{te}+E_{tm}=\frac{2 E_i \cos\theta_i}{\cos\theta_i+\cos\theta_t}.
\end{eqnarray}
We also have a reflected field which reflects off the surface in a specular fashion with an amplitude of,
\begin{eqnarray}
\label{eq:ReflAmp}
E_{is}=E_{ism}-E_{ise}=\frac{ E_i( \cos\theta_i-\cos\theta_t)}{\cos\theta_i+\cos\theta_t}. 
\end{eqnarray}
Note however, that this reflected field is much smaller than the refracted field. 
\\ \\
To further show that these are the only fields generated by combined impedance/admittance surface, we can look at the power entering and leaving the scattered surfaces by integrating the fields over an area enclosed by the surface as shown in Fig.~\ref{fig:PlaneWaveRefractionImpAdmSurf}. The power entering and leaving the surface is given by,
\begin{eqnarray}
P = \frac{1}{2} \iint_S \Re \left[ \mathbf{E}\times\mathbf{H}^{\ast}\right]\cdot\mathbf{\hat{n}}dS.
\end{eqnarray}
For plane-waves this reduces to the power in the normal component of the field that enters or leaves a surface with area, $A$, which gives,
\begin{eqnarray}
P = \frac{A}{\eta}(|E_i|^2\cos\theta_i-E_t^2\cos\theta_t-E_{is}^2\cos\theta_i)=0
\end{eqnarray}
which demonstrates that all the power from the incident plane-wave is scattered into the refracted and reflected beam.  
\\ \\
An equivalent setup for an impedance and admittance surface that reflects an incident plane-wave along an arbitrary angle can easily be extended from the derivation given here by simply negating the direction of $E_{tm}$ in Fig.~\ref{fig:PlaneWaveRefractionAdmSurf} and re-deriving the subsequent equations.
\begin{figure}[!h]
\centering
\includegraphics[clip=true, trim= 0cm 3.5cm 0cm 4.0cm, scale=0.3]{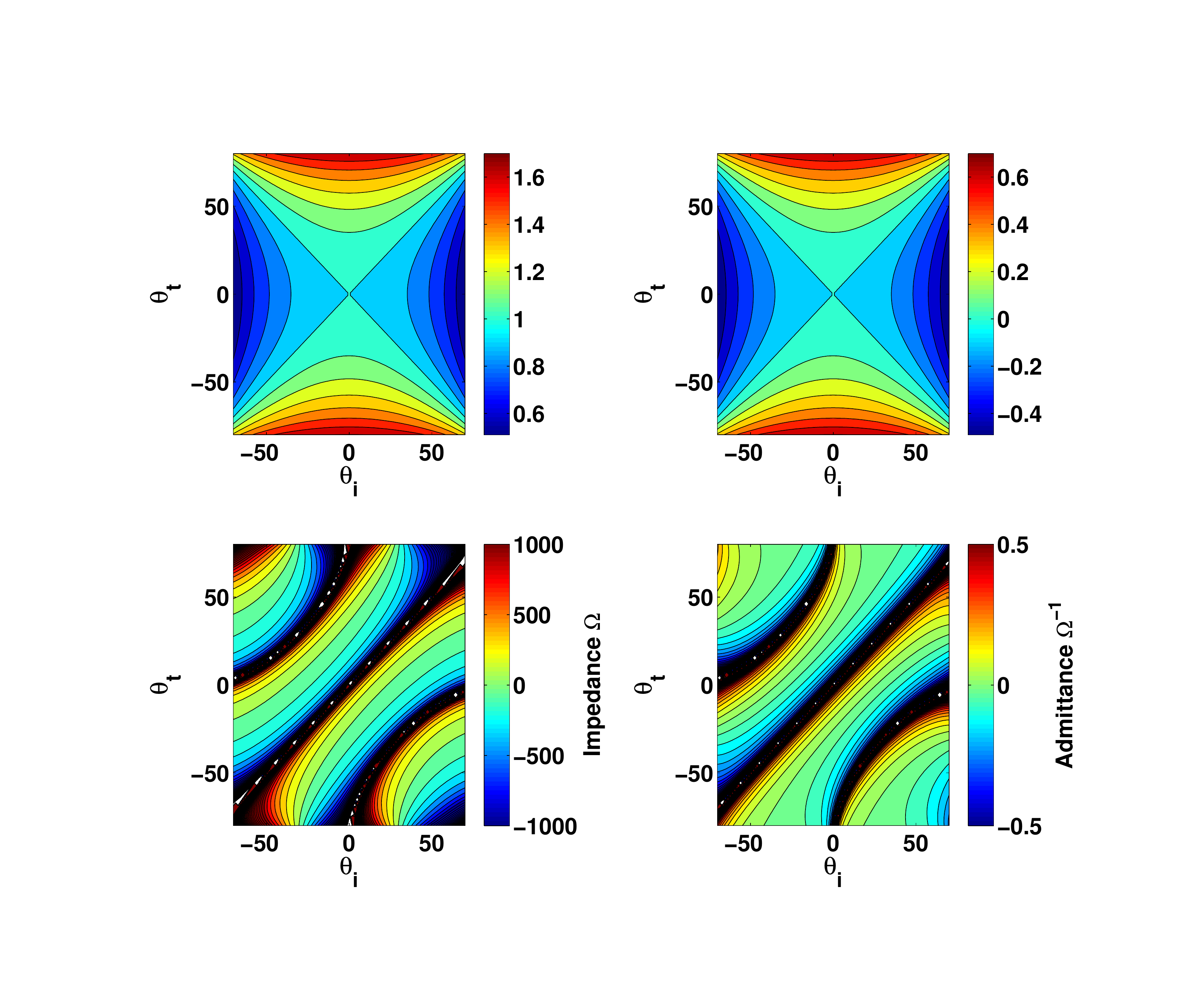}
\caption{Design curves summarizing refraction through an impedance and admittance surface for an incident plane wave of amplitude $E_i=1$. All plots are drawn with incident angle, $\theta_i$, on the horizontal axis and the refracted angle, $\theta_t$, on the vertical axis. The top-left plot gives us the amplitude of the refracted beam, $E_t$ while the top-right plot gives us the amplitude of the reflected beam $E_{is}$.  The bottom two curves are contour plots of the  impedance and admittance of the screen at a given point, ($y=\lambda$) for different incident and reflected angles.  }
\label{fig:SurfaceRefractionDesignCurves}
\end{figure}
\\ \\
We can summarize these results in a set of design curves which describe how a plane-wave refracts through a combined impedance and admittance surface.  This is plotted in Fig.~\ref{fig:SurfaceRefractionDesignCurves} and describes the solutions of $E_{t}$, $E_{is}$, $X_s$ and $B_s$ for all possible combinations of $\theta_{i}$ and $\theta_{t}$. The plots on the top-left and top-right of Fig.~\ref{fig:SurfaceRefractionDesignCurves} give us the amplitude of $E_t$ and $E_{is}$ respectively for different values of $\theta_t$ and $\theta_i$.  Note that for most reasonable angles of incidence the amplitude of the refracted beam, $E_{t}$ is close to $1$ while the reflected beam is close to $0$ showing that our combined surface is ideal for refraction with minimal reflection.  In the bottom-left and bottom-right plots we see contours of constant impedance and susceptance respectively for different combinations of $\theta_t$ and $\theta_i$. Note here, that for small values of both $\theta_t$ and $\theta_i$ the curves overlap. This can be seen from inserting small-angle expressions for the angles into Eq.~\ref{eq:Xs} and Eq.~\ref{eq:Bs}.  This implies that for small angles, a fixed combined impedance/admittance surface can predictably refract an incident plane wave of an arbitrary angle of incidence. However, as both these values diverge the impedance and admittance curves diverge from each other as well. In general then, like the active cloaks shown here and in \cite{Selvanayagam_Eleftheriades_2012_4}, we are constrained with knowing the incident field to design our surface for a desired angle of refraction, $\theta_t$. 
\\ \\
Again, these admittance and impedance surfaces act like a passive Huygens source which generates the required wavefronts to construct the refracted field. The incident plane-wave along a direction $\theta_i$ induces electric and magnetic currents on the impedance and admittance surface respectively which independently radiate a scattered field which interferes to form a refracted plane-wave in the direction of $\theta_t$
\\
\\
If we had not been so careful as to constrain the amplitudes of the refracted fields we would have arrived at impedance and admittance surfaces which had a non-zero and possible negative real part.  This would have implied a surface which scattered more power then was incident on it and would be analogous to the example in Section \ref{subsec:ActiveRefraction} where we used an active dipole array to refract a plane-wave. As we have shown for refraction, the power in the incident field can simply be redirected into the refracted field by a passive surface as long as the amplitudes are constrained. Thus an active surface is not truly needed (though it does demonstrate the concept of a field discontinuity). To contrast, for cloaking the power in the incident and scattered field is already conserved \cite{Born_Wolf}. Thus to enforce a discontinuity in the field which cancels the scattered field the power must come from elsewhere, in this case an array of electric and magnetic dipoles (Note that this is different than bending the light around the object as in \cite{pendry_etal_2006} or resonating out a specific multipole order of the scattered field \cite{Alu_Engheta_2008} which are passive cloaking schemes but which do not superimpose a field into the problem).
\\ \\
%While the development such surfaces are outside the scope of this work, the ability to synthesize such a surface would give new degrees of freedom in refracting and reflecting a plane-wave and could lead to other exotic behaviour.
Finally, we note that this construction of impedance and admittance surfaces gives the full description of how to refract a plane-wave across a surface.  As a point of comparison, we can look at how refraction and reflection of a plane-wave across a material interface is derived. Here we notice three attributes about this process. First it is derived from the boundary conditions at a material interface. Secondly, it tells us how the field will refract and reflect (Snell's Law). Third, it tells us the amplitudes of the field (the Fresnel reflection coefficients). Here we have the same attributes but realized using a combined impedance/admittance surface.  for a surface. Starting from the boundary conditions given by impedance and admittance surfaces, we have derived how the field will refract across a surface for different impedances (Eq.~\ref{eq:Xs} and Eq.~\ref{eq:Bs}) and we have found the amplitudes of those fields in Eq.~\ref{eq:RefractedAmp} and Eq.~\ref{eq:ReflAmp}. This gives us a complete picture for how to tailor a surface to refract (or reflect) an incident plane-wave. 
\\ \\
In comparison, the generalized refraction law in Eq.~\ref{eq:GeneralRefraction} simply tells us the phase gradient between the two refracted plane-waves.  It does not tell us how to synthesize a surface.  (And as shown, trying to synthesize a surface which simply mimics the phase gradient is incomplete). Thus this equation is descriptive, as it describes how a plane-wave refracts across a surface, but it is not prescriptive in that it does not tell us how to design our surface.  Eq.~\ref{eq:GeneralRefraction} is useful as a preliminary step as we can envision an arbitrary gradient along a surface which encodes some functionality (focusing, beam steering, vortexes) with respect to some incident field. However we would then need to use the equations given in Eq.~\ref{eq:Xs} and Eq.~\ref{eq:Bs} to synthesize the required surface using both electric and magnetic currents generated by the impedance and admittance surfaces.

\subsection{Examples}
We can verify these results by using two simple full-wave simulations.  The first example is to use the two-dimensional method of moments to solve for the required impedance and admittance as well as the fields radiated by the surface. The second example  is using a commercial full-wave solver (HFSS) to demonstrate a very basic physical implementation of this concept above.

\subsubsection{Method of Moments Verification}
To analyze the refraction through a surface we can use a two-dimensional method of moments procedure to numerically synthesize the required admittance and impedance as well as to find the radiation from the induced currents on the surface, confirming the results above \cite{Grbic_Merlin_2008,Gibson}.
\\ \\
For the impedance surface the induced electric currents on the surface can be found from examining the relationship between the incident, scattered and total fields at the boundary
\begin{eqnarray}
\mathbf{E_{tot}}|_{x=0}&=&\mathbf{E_i}|_{x=0}+\mathbf{E_s}|_{x=0}.
\end{eqnarray}
The scattered field is created by the radiation of the electric currents induced on the surface which is given by \cite{Gibson},
\begin{eqnarray}
\label{eq:JsMom}
\mathbf{E_{tot}}|_{x=0}&=E_i e^{-jky\sin\theta_i}\mathbf{\hat{z}} - \nonumber \\ \frac{\omega\mu}{4} \int_{-L/2}^{L/2} \mathbf{J_s}(y') H_o^2(k|y-y'|)dy',
\end{eqnarray}
where $L$ is the length of the impedance surface.
\\ \\
Likewise, for the admittance surface, we can solve for the magnetic currents by examining the incident, total and scattered field at the boundary
\begin{eqnarray}
\mathbf{\hat{n}}\times\mathbf{H_{tot}}|_{x=0}&=&\mathbf{\hat{n}}\times\mathbf{H_i}|_{x=0}+\mathbf{\hat{n}}\times\mathbf{H_s}|_{x=0}.
\end{eqnarray}
Here we must find the tangential magnetic field created by the radiation of the magnetic currents on the surface which is given to be,
\begin{eqnarray}
\label{eq:MsMom}
\mathbf{\hat{n}}\times\mathbf{H_{tot}}|_{x=0}=\frac{E_i}{\eta}\cos\theta_i e^{-jky\sin\theta_i}\mathbf{\hat{x}} \nonumber \\ -\frac{1}{4\omega\mu}\left(k^2+\frac{\partial^2}{\partial y^2}\right) \int_{-L/2}^{L/2} \mathbf{M_s}(y') H_o^2(k|y-y'|)dy',
\end{eqnarray}
Both Eq.~\ref{eq:JsMom} and Eq.~\ref{eq:MsMom} can be solved for $\mathbf{J_s}$ and $\mathbf{M_s}$ respectively using the method of moments since we know both the desired total fields and incident field at the boundary.  To solve, the impedance surface is discretized into N segments and is solved for using pulse-basis functions and point matching \cite{Gibson}. Once $\mathbf{J_s}$ and $\mathbf{M_s}$ are found, these can be respectively inserted into Eq.~\ref{eq:ZImpSurf} and Eq.\ref{eq:YAdmSurf}.
\begin{figure}[!h]
\centering
\includegraphics[clip=true, trim= 0cm 3cm 0cm 3cm, scale=0.25]{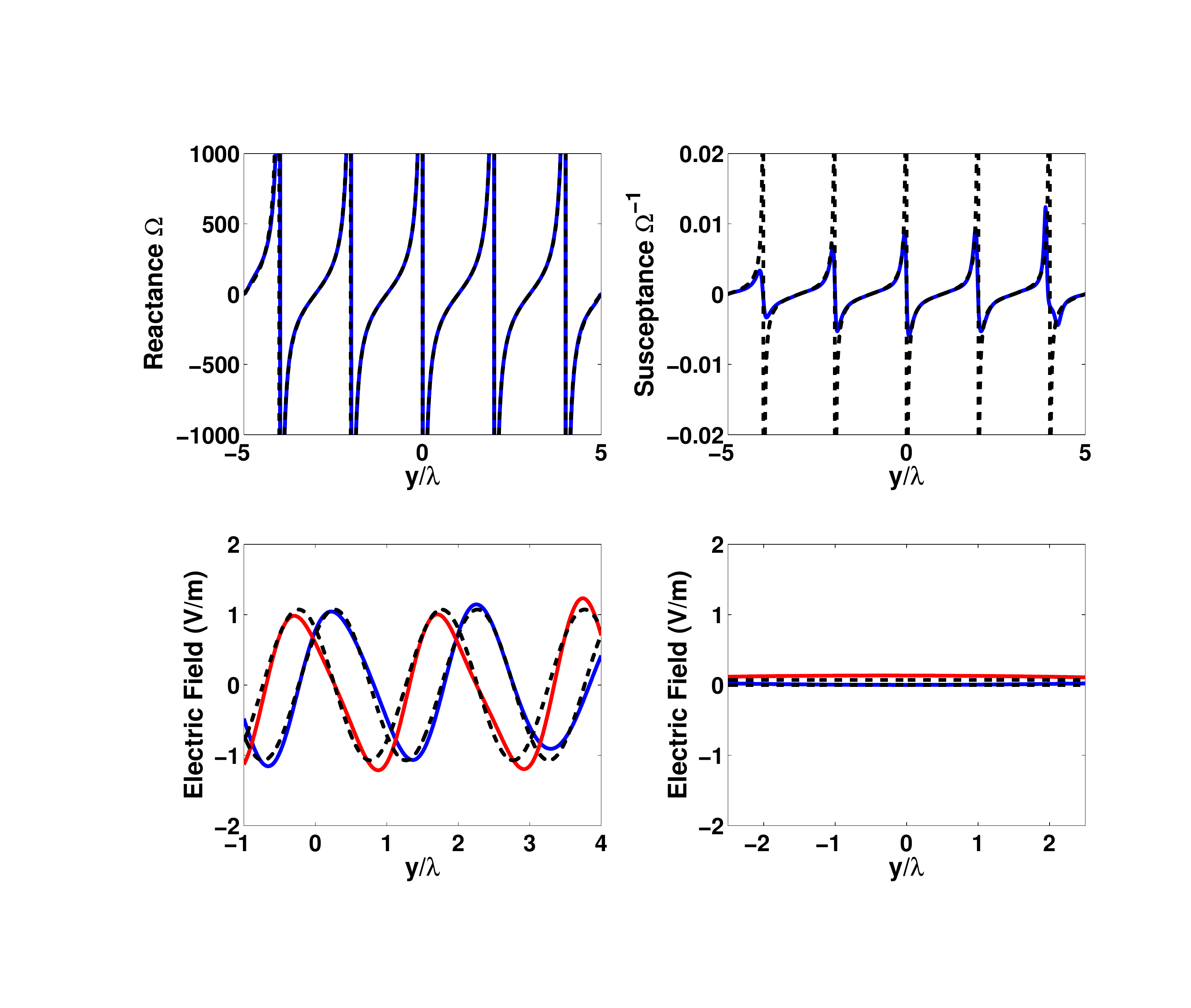}
\caption{Method of moments verification of an impedance and admittance surface to refract an incident plane-wave. Solid, coloured curves are calculated method of moments results.  Dashed black curves are theoretical results from the previous section. The top-left and top-right figures are the calculated reactance and susceptance of the two surfaces. We can see good agreement with the theoretical results as well as their non-linear dependence on the spatial coordinate of the surface.  The bottom-left and bottom-right plots show the calculated and theoretical electric field on side-two and side-one of the surface respectively (real and imaginary parts in red and blue respectively ). Note the much larger amplitude of the the refracted beam compared to the reflected beam. }
\label{fig:MOMrefractionImpedanceRadiation}
\end{figure}
\\ \\

To demonstrate this we take an example where our incident field is in the normal direction, $\theta_i=0^{\circ}$ and our desired refracted beam is along $\theta_t=30^{\circ}$. Our impedance and admittance surfaces are $L=10\lambda$ long at $f=1.5$~GHz. We divide our surface up into $N=1000$ segments that are $\lambda/100$ long. The calculated reactance and susceptance are plotted in Fig.~\ref{fig:MOMrefractionImpedanceRadiation} along with the theoretical values given by Eq.~\ref{eq:Xs} and Eq.~\ref{eq:Bs} where good agreement can be seen between the two sets of curves. We can also find the total field on either side of the screen by finding the fields radiated from both sets of currents. On side two of the boundary these fields must be summed with the incident field. These radiated fields are given by,
\begin{eqnarray}
\mathbf{E_{s,elec}}&=&-\frac{\omega\mu}{4} \int_{-L/2}^{L/2} \mathbf{J_s}(y') H_o^2(k|\sqrt{(y-y')^2+z^2}|)dy',\\
\mathbf{E_{s,mag}}&=&-\frac{j}{4} \int_{-L/2}^{L/2} \nabla\times \Big[ \mathbf{M_s}(y') H_o^2(k|\sqrt{(y-y')^2+z^2}|)\Big] dy',\\
\mathbf{E_s}&=&\mathbf{E_{s,elec}}+\mathbf{E_{s,mag}}
\end{eqnarray}
Plotting these radiated fields in Fig.~\ref{fig:MOMrefractionImpedanceRadiation}, we find a field that resembles the theoretical prediction fairly well on either side of the screen with the discrepancies coming from the finite nature of the screen. On side two of the screen we have a field  which resembles a plane-wave propagating along $\theta_t=30^{\circ}$. On side one of the screen there is a small reflected field as predicted by Eq.~\ref{eq:ReflAmp} which is plane-wave along the $\theta_i=0^{\circ}$ direction. Note however that the amplitude of these fields is much smaller than the refracted fields on the other side of the screen. This indicates that a surface of induced magnetic and electric currents through impedances and admittances allows for a discontinuity in the electric and magnetic field to be realized.

\subsubsection{Physical Implementation}
\begin{figure*}[!h]
\centering
\includegraphics[clip=true, trim= 0cm 3.5cm 0cm 4.0cm, scale=0.4]{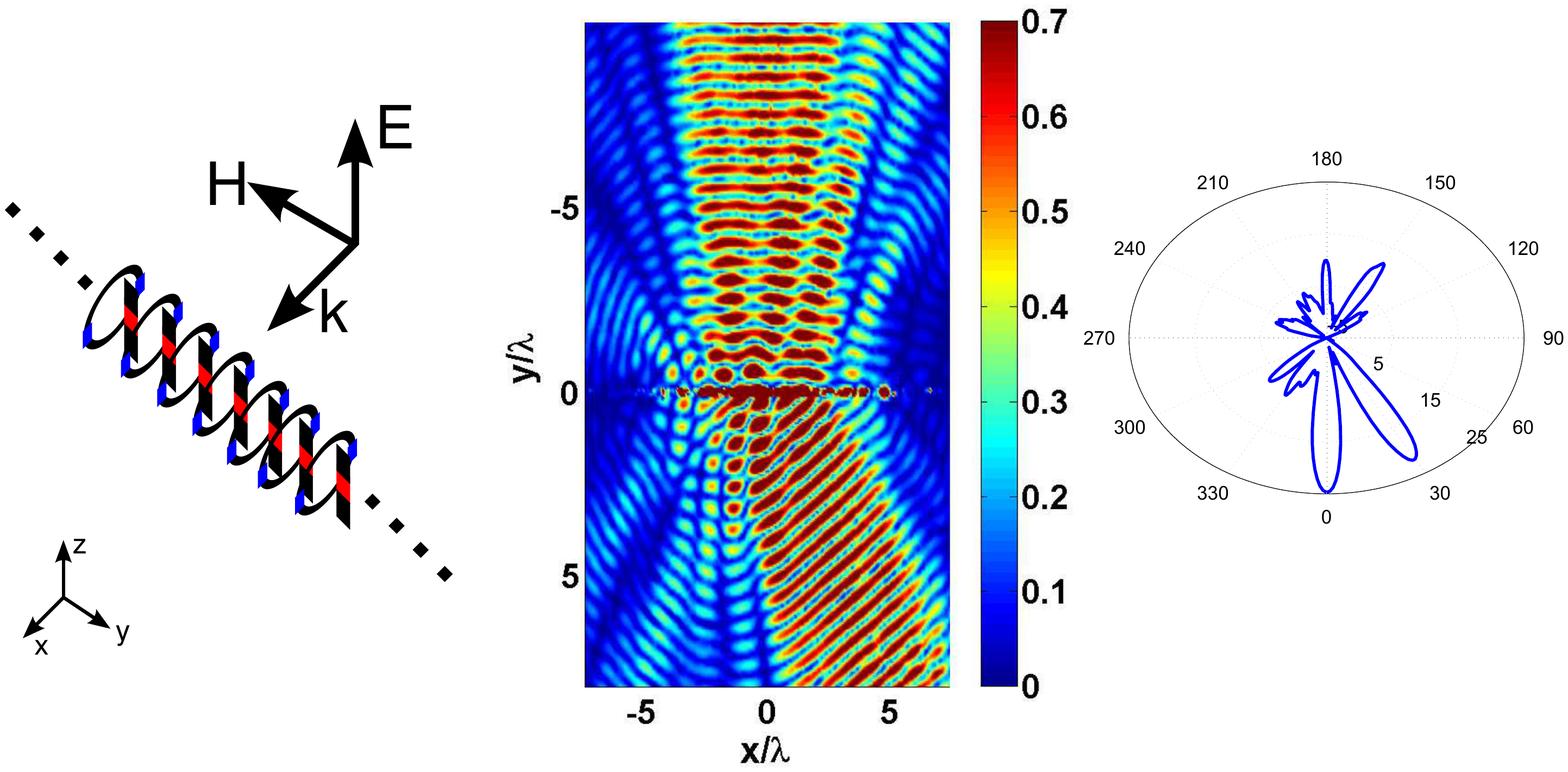}
\caption{Simulation results from HFSS for a physically realized impedance and admittance screen.  On the left a schematic of the surface. The impedance screen is made of reactively loaded dipoles with the reactive loading shown in red. The admittance screen is made of reactively loaded loops with the loading shown in blue. In the middle is a plot of the total vertical electric field for a simulation of the impedance screen for an incident Gaussian beam.  The plot on the right is a far-field plot of the scattered electric field. } %We can see the incident field refracts mainly into the refracted beam with some smaller spurious refractions and reflections. The plot on the right is a far-field plot of the scattered electric field. The majority of the scattered field is in a beam along $0^{\circ}$ and a beam along $30^{\circ}$. The beam along $0^{\circ}$ destructively interferes with the incident field as shown on the field plot on the right leaving the refracted beam only. As expected there is a small reflected field at $180^{\circ}$, while the reflected field is not completely canceled out at $150^{\circ}$. This can be improved upon further optimization of the screen.  }
\label{fig:HFSSGaussianBeamImpAdmEfieldCombined}
\end{figure*}

As stated before, the impedance and admittance surfaces set up induced electric and magnetic currents to impose a discontinuity in the field. To physically implement such a surface we can take a cue from the active version of this idea and use passive dipoles and loops to create our induced electric and magnetic currents. We will demonstrate this idea by using an array of loops loaded with a reactive impedance to create our induced magnetic current. We will use dipoles loaded reactively to create induced electric currents. By varying the reactive loading on each loop/dipole the impedance and admittance of the surface can be tuned.  
\\  \\
To model this we will use Ansys's HFSS.  Again we will use the same example as before at $1.5$~GHz with an incident field at $\theta_i=0^{\circ}$ and a refracted field at $\theta_t=30^{\circ}$. We will place our simulation in a parallel-plate waveguide to simplify the computation and emulate a 2-D environment. Taking the impedances and admittances given in Fig.~\ref{fig:MOMrefractionImpedanceRadiation}, each loop/dipole is designed to implement the desired impedance at its location along the surface, with the loops/dipoles spaced every $\lambda/10$. This is shown in Fig.~\ref{fig:HFSSGaussianBeamImpAdmEfieldCombined}. To illuminate the surface a Gaussian beam is used with a $3\lambda$ focal spot designed to occur at the surface. With this design we offer a couple of caveats. First the unit cells are not optimal and better designs are possible. Secondly the impedance variation along the surface is not optimized for a finite beam such as a Gaussian beam.  Nonetheless the main point can be demonstrated here with this simple array of passive loops and dipoles. 
\\ \\
The simulated results are shown in Fig.~\ref{fig:HFSSGaussianBeamImpAdmEfieldCombined} where the field-plot at 1.5GHz is shown. Here we can see that the Gaussian beam is refracted by the surface from an incident direction of $\theta_i=0^{\circ}$ to a refracted direction of $\theta_t=30^{\circ}$. Reflections are also minimal though any discrepancies can be attributed to the reasons given above. The far-field pattern of the scattered field is also plotted  in Fig.~\ref{fig:HFSSGaussianBeamImpAdmEfieldCombined} where we can see the main refracted beam at $\theta=30^{\circ}$. The peak at $\theta=0^{\circ}$ is the field radiated by the surface to cancel out the incident field on side two of the surface (as we are plotting the scattered field). There is a reflected peak at $\theta=180^{\circ}$ due to the small reflected field that is generated by the surface and a reflected peak at $\theta=150^{\circ}$ that is not completely canceled out due to the imperfections in the surface. This can be improved upon further optimization of the screen. 

%\subsection{Observations}

\section{Conclusion}
Through the examples presented in this work, we have demonstrated how the idea of discontinuous fields can be used to alter, shape and control electromagnetic wavefronts. By placing electric and magnetic currents at a boundary, the fields can be altered in a variety of novel ways including the ability to cloak an object and the ability to refract (or reflect) an incident field.  We have shown that we can create our electric and magnetic currents in one of two main ways, either by placing a discrete array of active electric and magnetic dipoles which impress a discontinuity in the field. Or by using a passive impedance and admittance surface, on which electric and magnetic currents are induced to create a discontinuity.
\\ \\
This approach of imposing a discontinuity in the electromagnetic field opens up many new possibilities across the electromagnetic spectrum. Some interesting areas include the synthesis of active electric and magnetic dipole arrays as well as passive impedance and admittance surfaces at both microwave frequencies, using well known technologies in phased-arrays and impedance surface \cite{Hansen,Sievenpiper_etal_1999}, and at optical and THz frequencies using more recent developments such as nano-antennas, nanophotonic phased-ararys and nano-circuits \cite{Engheta_2007,Novotny_Hulst_2011,Sun_etal_2013} . The design of novel devices using discontinuous fields including lenses, cloaks, reflectors, can also be envisioned using this new technique and presents a promising avenue for future work. 

%finiteness

%%%%%%%%%%%%%%%%%%%%%%% References %%%%%%%%%%%%%%%%%%%%%%%%%
%\begin{thebibliography}{99}
%\bibliographystyle{osajnl}
%\bibliography{../../bibtex/thesis}

\begin{thebibliography}{25}%
\makeatletter
\providecommand \@ifxundefined [1]{%
 \@ifx{#1\undefined}
}%
\providecommand \@ifnum [1]{%
 \ifnum #1\expandafter \@firstoftwo
 \else \expandafter \@secondoftwo
 \fi
}%
\providecommand \@ifx [1]{%
 \ifx #1\expandafter \@firstoftwo
 \else \expandafter \@secondoftwo
 \fi
}%
\providecommand \natexlab [1]{#1}%
\providecommand \enquote  [1]{``#1''}%
\providecommand \bibnamefont  [1]{#1}%
\providecommand \bibfnamefont [1]{#1}%
\providecommand \citenamefont [1]{#1}%
\providecommand \href@noop [0]{\@secondoftwo}%
\providecommand \href [0]{\begingroup \@sanitize@url \@href}%
\providecommand \@href[1]{\@@startlink{#1}\@@href}%
\providecommand \@@href[1]{\endgroup#1\@@endlink}%
\providecommand \@sanitize@url [0]{\catcode `\\12\catcode `\$12\catcode
  `\&12\catcode `\#12\catcode `\^12\catcode `\_12\catcode `\%12\relax}%
\providecommand \@@startlink[1]{}%
\providecommand \@@endlink[0]{}%
\providecommand \url  [0]{\begingroup\@sanitize@url \@url }%
\providecommand \@url [1]{\endgroup\@href {#1}{\urlprefix }}%
\providecommand \urlprefix  [0]{URL }%
\providecommand \Eprint [0]{\href }%
\providecommand \doibase [0]{http://dx.doi.org/}%
\providecommand \selectlanguage [0]{\@gobble}%
\providecommand \bibinfo  [0]{\@secondoftwo}%
\providecommand \bibfield  [0]{\@secondoftwo}%
\providecommand \translation [1]{[#1]}%
\providecommand \BibitemOpen [0]{}%
\providecommand \bibitemStop [0]{}%
\providecommand \bibitemNoStop [0]{.\EOS\space}%
\providecommand \EOS [0]{\spacefactor3000\relax}%
\providecommand \BibitemShut  [1]{\csname bibitem#1\endcsname}%
\let\auto@bib@innerbib\@empty
%</preamble>
\bibitem [{\citenamefont {Selvanayagam}\ and\ \citenamefont
  {Eleftheriades}(2012)}]{Selvanayagam_Eleftheriades_2012_4}%
  \BibitemOpen
  \bibfield  {author} {\bibinfo {author} {\bibfnamefont {M.}~\bibnamefont
  {Selvanayagam}}\ and\ \bibinfo {author} {\bibfnamefont {G.~V.}\ \bibnamefont
  {Eleftheriades}},\ }\href@noop {} {\bibfield  {journal} {\bibinfo  {journal}
  {IEEE Antennas and Wireless Propagation Letters}\ } (\bibinfo {year}
  {2012})}\BibitemShut {NoStop}%
\bibitem [{\citenamefont {Harrington}(2001)}]{Harrington}%
  \BibitemOpen
  \bibfield  {author} {\bibinfo {author} {\bibfnamefont {R.~F.}\ \bibnamefont
  {Harrington}},\ }\href@noop {} {\emph {\bibinfo {title} {Time-Harmonic
  Electromagnetic Fields}}}\ (\bibinfo  {publisher} {Wiley-Interscience},\
  \bibinfo {year} {2001})\BibitemShut {NoStop}%
\bibitem [{\citenamefont {Hansen}(2009)}]{Hansen}%
  \BibitemOpen
  \bibfield  {author} {\bibinfo {author} {\bibfnamefont {R.~C.}\ \bibnamefont
  {Hansen}},\ }\href@noop {} {\emph {\bibinfo {title} {Phased Array
  Antennas}}}\ (\bibinfo  {publisher} {Wiley-Interscience},\ \bibinfo {year}
  {2009})\BibitemShut {NoStop}%
\bibitem [{\citenamefont {Kwon}\ and\ \citenamefont
  {Pozar}(2009)}]{Kwon_Pozar_2009}%
  \BibitemOpen
  \bibfield  {author} {\bibinfo {author} {\bibfnamefont {D.-H.}\ \bibnamefont
  {Kwon}}\ and\ \bibinfo {author} {\bibfnamefont {D.~M.}\ \bibnamefont
  {Pozar}},\ }\href {\doibase 10.1109/TAP.2009.2025975} {\bibfield  {journal}
  {\bibinfo  {journal} {Antennas and Propagation, IEEE Transactions on}\
  }\textbf {\bibinfo {volume} {57}},\ \bibinfo {pages} {3720} (\bibinfo {year}
  {2009})}\BibitemShut {NoStop}%
\bibitem [{\citenamefont {Oppenheim}\ \emph {et~al.}(1997)\citenamefont
  {Oppenheim}, \citenamefont {Willsky},\ and\ \citenamefont
  {Nawab}}]{Oppenheim}%
  \BibitemOpen
  \bibfield  {author} {\bibinfo {author} {\bibfnamefont {A.~V.}\ \bibnamefont
  {Oppenheim}}, \bibinfo {author} {\bibfnamefont {A.~S.}\ \bibnamefont
  {Willsky}}, \ and\ \bibinfo {author} {\bibfnamefont {S.~H.}\ \bibnamefont
  {Nawab}},\ }\href@noop {} {\emph {\bibinfo {title} {Signals and Systems}}}\
  (\bibinfo  {publisher} {Prentice Hall},\ \bibinfo {year} {1997})\BibitemShut
  {NoStop}%
\bibitem [{\citenamefont {Tretyakov}(2003)}]{Tretyakov}%
  \BibitemOpen
  \bibfield  {author} {\bibinfo {author} {\bibfnamefont {S.}~\bibnamefont
  {Tretyakov}},\ }\href@noop {} {\emph {\bibinfo {title} {Analytical Modeling
  in Applied Electromagnets}}}\ (\bibinfo  {publisher} {Artech House},\
  \bibinfo {year} {2003})\BibitemShut {NoStop}%
\bibitem [{\citenamefont {Grbic}\ and\ \citenamefont
  {Merlin}(2008)}]{Grbic_Merlin_2008}%
  \BibitemOpen
  \bibfield  {author} {\bibinfo {author} {\bibfnamefont {A.}~\bibnamefont
  {Grbic}}\ and\ \bibinfo {author} {\bibfnamefont {R.}~\bibnamefont {Merlin}},\
  }\href {\doibase 10.1109/TAP.2008.929436} {\bibfield  {journal} {\bibinfo
  {journal} {Antennas and Propagation, IEEE Transactions on}\ }\textbf
  {\bibinfo {volume} {56}},\ \bibinfo {pages} {3159} (\bibinfo {year}
  {2008})}\BibitemShut {NoStop}%
\bibitem [{\citenamefont {Chen}\ \emph {et~al.}(2008)\citenamefont {Chen},
  \citenamefont {Luo}, \citenamefont {Ma},\ and\ \citenamefont
  {Chan}}]{Chen_etal_2008}%
  \BibitemOpen
  \bibfield  {author} {\bibinfo {author} {\bibfnamefont {H.}~\bibnamefont
  {Chen}}, \bibinfo {author} {\bibfnamefont {X.}~\bibnamefont {Luo}}, \bibinfo
  {author} {\bibfnamefont {H.}~\bibnamefont {Ma}}, \ and\ \bibinfo {author}
  {\bibfnamefont {C.}~\bibnamefont {Chan}},\ }\href {\doibase
  10.1364/OE.16.014603} {\bibfield  {journal} {\bibinfo  {journal} {Opt.
  Express}\ }\textbf {\bibinfo {volume} {16}},\ \bibinfo {pages} {14603}
  (\bibinfo {year} {2008})}\BibitemShut {NoStop}%
\bibitem [{\citenamefont {Gallina}\ \emph {et~al.}(2010)\citenamefont
  {Gallina}, \citenamefont {Castaldi}, \citenamefont {Galdi}, \citenamefont
  {Al\`u},\ and\ \citenamefont {Engheta}}]{gallina_engheta_2010}%
  \BibitemOpen
  \bibfield  {author} {\bibinfo {author} {\bibfnamefont {I.}~\bibnamefont
  {Gallina}}, \bibinfo {author} {\bibfnamefont {G.}~\bibnamefont {Castaldi}},
  \bibinfo {author} {\bibfnamefont {V.}~\bibnamefont {Galdi}}, \bibinfo
  {author} {\bibfnamefont {A.}~\bibnamefont {Al\`u}}, \ and\ \bibinfo {author}
  {\bibfnamefont {N.}~\bibnamefont {Engheta}},\ }\href {\doibase
  10.1103/PhysRevB.81.125124} {\bibfield  {journal} {\bibinfo  {journal} {Phys.
  Rev. B}\ }\textbf {\bibinfo {volume} {81}},\ \bibinfo {pages} {125124}
  (\bibinfo {year} {2010})}\BibitemShut {NoStop}%
\bibitem [{\citenamefont {Gibson}(2008)}]{Gibson}%
  \BibitemOpen
  \bibfield  {author} {\bibinfo {author} {\bibfnamefont {W.~C.}\ \bibnamefont
  {Gibson}},\ }\href@noop {} {\emph {\bibinfo {title} {The Method of Moments in
  Electromagnetics}}}\ (\bibinfo  {publisher} {Chapman and Hall/CRC},\ \bibinfo
  {year} {2008})\BibitemShut {NoStop}%
\bibitem [{\citenamefont {Miller}(2006)}]{Miller_2006}%
  \BibitemOpen
  \bibfield  {author} {\bibinfo {author} {\bibfnamefont {D.~A.}\ \bibnamefont
  {Miller}},\ }\href@noop {} {\bibfield  {journal} {\bibinfo  {journal} {Optics
  Express}\ }\textbf {\bibinfo {volume} {14}} (\bibinfo {year}
  {2006})}\BibitemShut {NoStop}%
\bibitem [{\citenamefont {Ryan}\ \emph {et~al.}(2010)\citenamefont {Ryan},
  \citenamefont {Chaharmir}, \citenamefont {Shaker}, \citenamefont {Bray},
  \citenamefont {Antar},\ and\ \citenamefont {Ittipiboon}}]{Ryan_etal_2010}%
  \BibitemOpen
  \bibfield  {author} {\bibinfo {author} {\bibfnamefont {C.~G.~M.}\
  \bibnamefont {Ryan}}, \bibinfo {author} {\bibfnamefont {M.}~\bibnamefont
  {Chaharmir}}, \bibinfo {author} {\bibfnamefont {J.}~\bibnamefont {Shaker}},
  \bibinfo {author} {\bibfnamefont {J.}~\bibnamefont {Bray}}, \bibinfo {author}
  {\bibfnamefont {Y.~M.~M.}\ \bibnamefont {Antar}}, \ and\ \bibinfo {author}
  {\bibfnamefont {A.}~\bibnamefont {Ittipiboon}},\ }\href {\doibase
  10.1109/TAP.2010.2044356} {\bibfield  {journal} {\bibinfo  {journal}
  {Antennas and Propagation, IEEE Transactions on}\ }\textbf {\bibinfo {volume}
  {58}},\ \bibinfo {pages} {1486} (\bibinfo {year} {2010})}\BibitemShut
  {NoStop}%
\bibitem [{\citenamefont {Lau}\ and\ \citenamefont {Hum}(2012)}]{Lau_Hum_2012}%
  \BibitemOpen
  \bibfield  {author} {\bibinfo {author} {\bibfnamefont {J.}~\bibnamefont
  {Lau}}\ and\ \bibinfo {author} {\bibfnamefont {S.}~\bibnamefont {Hum}},\
  }\href {\doibase 10.1109/TAP.2012.2213054} {\bibfield  {journal} {\bibinfo
  {journal} {Antennas and Propagation, IEEE Transactions on}\ }\textbf
  {\bibinfo {volume} {60}},\ \bibinfo {pages} {5679 } (\bibinfo {year}
  {2012})}\BibitemShut {NoStop}%
\bibitem [{\citenamefont {Yu}\ \emph {et~al.}(2011)\citenamefont {Yu},
  \citenamefont {Genevet}, \citenamefont {Kats}, \citenamefont {Aieta},
  \citenamefont {Tetienne}, \citenamefont {Capasso},\ and\ \citenamefont
  {Gaburro}}]{Yu_etal_2011}%
  \BibitemOpen
  \bibfield  {author} {\bibinfo {author} {\bibfnamefont {N.}~\bibnamefont
  {Yu}}, \bibinfo {author} {\bibfnamefont {P.}~\bibnamefont {Genevet}},
  \bibinfo {author} {\bibfnamefont {M.~A.}\ \bibnamefont {Kats}}, \bibinfo
  {author} {\bibfnamefont {F.}~\bibnamefont {Aieta}}, \bibinfo {author}
  {\bibfnamefont {J.-P.}\ \bibnamefont {Tetienne}}, \bibinfo {author}
  {\bibfnamefont {F.}~\bibnamefont {Capasso}}, \ and\ \bibinfo {author}
  {\bibfnamefont {Z.}~\bibnamefont {Gaburro}},\ }\href {\doibase
  10.1126/science.1210713} {\bibfield  {journal} {\bibinfo  {journal}
  {Science}\ }\textbf {\bibinfo {volume} {334}},\ \bibinfo {pages} {333}
  (\bibinfo {year} {2011})},\ \Eprint
  {http://arxiv.org/abs/http://www.sciencemag.org/content/334/6054/333.full.pdf}
  {http://www.sciencemag.org/content/334/6054/333.full.pdf} \BibitemShut
  {NoStop}%
\bibitem [{\citenamefont {Kildishev}\ \emph {et~al.}(2013)\citenamefont
  {Kildishev}, \citenamefont {Boltasseva},\ and\ \citenamefont
  {Shalaev}}]{Kildishev_etal_2013}%
  \BibitemOpen
  \bibfield  {author} {\bibinfo {author} {\bibfnamefont {A.~V.}\ \bibnamefont
  {Kildishev}}, \bibinfo {author} {\bibfnamefont {A.}~\bibnamefont
  {Boltasseva}}, \ and\ \bibinfo {author} {\bibfnamefont {V.~M.}\ \bibnamefont
  {Shalaev}},\ }\href {\doibase 10.1126/science.1232009} {\bibfield  {journal}
  {\bibinfo  {journal} {Science}\ }\textbf {\bibinfo {volume} {339}} (\bibinfo
  {year} {2013}),\ 10.1126/science.1232009}\BibitemShut {NoStop}%
\bibitem [{\citenamefont {Tsai}\ \emph {et~al.}(2011)\citenamefont {Tsai},
  \citenamefont {Larouche}, \citenamefont {Tyler}, \citenamefont {Lipworth},
  \citenamefont {Jokerst},\ and\ \citenamefont {Smith}}]{Tsai_etal_2011}%
  \BibitemOpen
  \bibfield  {author} {\bibinfo {author} {\bibfnamefont {Y.-J.}\ \bibnamefont
  {Tsai}}, \bibinfo {author} {\bibfnamefont {S.}~\bibnamefont {Larouche}},
  \bibinfo {author} {\bibfnamefont {T.}~\bibnamefont {Tyler}}, \bibinfo
  {author} {\bibfnamefont {G.}~\bibnamefont {Lipworth}}, \bibinfo {author}
  {\bibfnamefont {N.~M.}\ \bibnamefont {Jokerst}}, \ and\ \bibinfo {author}
  {\bibfnamefont {D.~R.}\ \bibnamefont {Smith}},\ }\href {\doibase
  10.1364/OE.19.024411} {\bibfield  {journal} {\bibinfo  {journal} {Opt.
  Express}\ }\textbf {\bibinfo {volume} {19}},\ \bibinfo {pages} {24411}
  (\bibinfo {year} {2011})}\BibitemShut {NoStop}%
\bibitem [{\citenamefont {Steyskal}\ \emph {et~al.}(1979)\citenamefont
  {Steyskal}, \citenamefont {Hessel},\ and\ \citenamefont
  {Shmoys}}]{Steyskal_etal_1979}%
  \BibitemOpen
  \bibfield  {author} {\bibinfo {author} {\bibfnamefont {H.}~\bibnamefont
  {Steyskal}}, \bibinfo {author} {\bibfnamefont {A.}~\bibnamefont {Hessel}}, \
  and\ \bibinfo {author} {\bibfnamefont {J.}~\bibnamefont {Shmoys}},\
  }\href@noop {} {\bibfield  {journal} {\bibinfo  {journal} {IEEE Trans. on
  Antennas and Propagation}\ }\textbf {\bibinfo {volume} {AP-27}},\ \bibinfo
  {pages} {825} (\bibinfo {year} {1979})}\BibitemShut {NoStop}%
\bibitem [{\citenamefont {Balanis}(2005)}]{balanis}%
  \BibitemOpen
  \bibfield  {author} {\bibinfo {author} {\bibfnamefont {C.~A.}\ \bibnamefont
  {Balanis}},\ }\href@noop {} {\emph {\bibinfo {title} {Antenna Theory:
  Analysis and Design}}}\ (\bibinfo  {publisher} {Wiley-Interscience},\
  \bibinfo {year} {2005})\BibitemShut {NoStop}%
\bibitem [{\citenamefont {Born}\ and\ \citenamefont {Wolf}(1999)}]{Born_Wolf}%
  \BibitemOpen
  \bibfield  {author} {\bibinfo {author} {\bibfnamefont {M.}~\bibnamefont
  {Born}}\ and\ \bibinfo {author} {\bibfnamefont {E.}~\bibnamefont {Wolf}},\
  }\href@noop {} {\emph {\bibinfo {title} {Principle of Optics}}},\ \bibinfo
  {edition} {7th}\ ed.\ (\bibinfo  {publisher} {Cambridge University Press},\
  \bibinfo {year} {1999})\BibitemShut {NoStop}%
\bibitem [{\citenamefont {Schurig}\ \emph {et~al.}(2006)\citenamefont
  {Schurig}, \citenamefont {Pendry},\ and\ \citenamefont
  {Smith}}]{pendry_etal_2006}%
  \BibitemOpen
  \bibfield  {author} {\bibinfo {author} {\bibfnamefont {D.}~\bibnamefont
  {Schurig}}, \bibinfo {author} {\bibfnamefont {J.~B.}\ \bibnamefont {Pendry}},
  \ and\ \bibinfo {author} {\bibfnamefont {D.~R.}\ \bibnamefont {Smith}},\
  }\href@noop {} {\bibfield  {journal} {\bibinfo  {journal} {Opt. Express}\
  }\textbf {\bibinfo {volume} {14}},\ \bibinfo {pages} {9794} (\bibinfo {year}
  {2006})}\BibitemShut {NoStop}%
\bibitem [{\citenamefont {Al\'u}\ and\ \citenamefont
  {Engheta}(2008)}]{Alu_Engheta_2008}%
  \BibitemOpen
  \bibfield  {author} {\bibinfo {author} {\bibfnamefont {A.}~\bibnamefont
  {Al\'u}}\ and\ \bibinfo {author} {\bibfnamefont {N.}~\bibnamefont
  {Engheta}},\ }\href@noop {} {\bibfield  {journal} {\bibinfo  {journal}
  {Journal of Optics A: Pure and Applied Optics}\ }\textbf {\bibinfo {volume}
  {10}},\ \bibinfo {pages} {093002} (\bibinfo {year} {2008})}\BibitemShut
  {NoStop}%
\bibitem [{\citenamefont {Sievenpiper}\ \emph {et~al.}(1999)\citenamefont
  {Sievenpiper}, \citenamefont {Zhang}, \citenamefont {Broas}, \citenamefont
  {Alexopolous},\ and\ \citenamefont {Yablonovitch}}]{Sievenpiper_etal_1999}%
  \BibitemOpen
  \bibfield  {author} {\bibinfo {author} {\bibfnamefont {D.}~\bibnamefont
  {Sievenpiper}}, \bibinfo {author} {\bibfnamefont {L.}~\bibnamefont {Zhang}},
  \bibinfo {author} {\bibfnamefont {R.}~\bibnamefont {Broas}}, \bibinfo
  {author} {\bibfnamefont {N.}~\bibnamefont {Alexopolous}}, \ and\ \bibinfo
  {author} {\bibfnamefont {E.}~\bibnamefont {Yablonovitch}},\ }\href@noop {}
  {\bibfield  {journal} {\bibinfo  {journal} {IEEE Transactions on Microwave
  Theory and Techniques}\ }\textbf {\bibinfo {volume} {47}},\ \bibinfo {pages}
  {2059} (\bibinfo {year} {1999})}\BibitemShut {NoStop}%
\bibitem [{\citenamefont {Engheta}(2007)}]{Engheta_2007}%
  \BibitemOpen
  \bibfield  {author} {\bibinfo {author} {\bibfnamefont {N.}~\bibnamefont
  {Engheta}},\ }\href@noop {} {\bibfield  {journal} {\bibinfo  {journal}
  {Science}\ }\textbf {\bibinfo {volume} {317}},\ \bibinfo {pages} {1698}
  (\bibinfo {year} {2007})},\ \Eprint
  {http://arxiv.org/abs/http://www.sciencemag.org/content/317/5845/1698.full.pdf}
  {http://www.sciencemag.org/content/317/5845/1698.full.pdf} \BibitemShut
  {NoStop}%
\bibitem [{\citenamefont {Novotny}\ and\ \citenamefont {van
  Hulst}(2011)}]{Novotny_Hulst_2011}%
  \BibitemOpen
  \bibfield  {author} {\bibinfo {author} {\bibfnamefont {L.}~\bibnamefont
  {Novotny}}\ and\ \bibinfo {author} {\bibfnamefont {N.}~\bibnamefont {van
  Hulst}},\ }\href@noop {} {\bibfield  {journal} {\bibinfo  {journal} {Nature
  Photonics}\ }\textbf {\bibinfo {volume} {5}},\ \bibinfo {pages} {83}
  (\bibinfo {year} {2011})}\BibitemShut {NoStop}%
\bibitem [{\citenamefont {Sun}\ \emph {et~al.}(2013)\citenamefont {Sun},
  \citenamefont {Timurdogan}, \citenamefont {Yaacobi}, \citenamefont
  {Hosseini},\ and\ \citenamefont {Watts}}]{Sun_etal_2013}%
  \BibitemOpen
  \bibfield  {author} {\bibinfo {author} {\bibfnamefont {J.}~\bibnamefont
  {Sun}}, \bibinfo {author} {\bibfnamefont {E.}~\bibnamefont {Timurdogan}},
  \bibinfo {author} {\bibfnamefont {A.}~\bibnamefont {Yaacobi}}, \bibinfo
  {author} {\bibfnamefont {E.~S.}\ \bibnamefont {Hosseini}}, \ and\ \bibinfo
  {author} {\bibfnamefont {M.~R.}\ \bibnamefont {Watts}},\ }\href@noop {}
  {\bibfield  {journal} {\bibinfo  {journal} {Nature}\ }\textbf {\bibinfo
  {volume} {493}},\ \bibinfo {pages} {195} (\bibinfo {year}
  {2013})}\BibitemShut {NoStop}%
\end{thebibliography}
%\end{thebibliography}

%

\end{document}